\documentclass[superscriptaddress,aps,prb,amsmath,amssymb,notitlepage, ,twocolumn]{revtex4-1}

\usepackage[pdftex]{graphicx}
\usepackage{subfigure}
\usepackage{dcolumn}
\usepackage{bm}
\usepackage[usenames,dvipsnames,svgnames]{xcolor}
\usepackage[colorlinks]{hyperref}
\hypersetup{
    linkcolor=BrickRed,
    urlcolor=blue,
    citecolor=MidnightBlue,
}

\newcommand{\bra}[1]{\langle #1 \vert}
\newcommand{\ket}[1]{\vert #1 \rangle}
\newcommand{\braket}[1]{\langle #1 \rangle}

\renewcommand{\vec}[1]{{\mathbf #1}}
\newcommand{\dvec}[1]{{\hat{\bm #1}}}
\newcommand{\mat}[1]{{\hat{\mathbf #1}}}

\newcommand{\STANECE}[0]{\affiliation{Electrical Engineering, Stanford University, Stanford, California 94305, USA}}
\newcommand{\UIUCECE}[0]{\affiliation{Department of Electrical and Computer Engineering, University of Illinois at Urbana-Champaign, Urbana, IL 61801, USA}}
\newcommand{\UIUCPHYS}[0]{\affiliation{Department of Physics, University of Illinois at Urbana-Champaign, Urbana, IL 61801, USA}}
\newcommand{\UIUCMNTL}[0]{\affiliation{Micro and Nanotechnology Laboratory, University of Illinois, 208 N. Wright Street, Urbana IL 61801, USA}}
\newcommand{\UIUCMAT}[0]{\affiliation{Department of Materials Science and Engineering and Materials
Research Laboratory, University of Illinois, Urbana, Illinois 61801, USA}}
\bibpunct{(}{)}{;}{s}{,}{,}

\begin{document}

\title{Impact of thermal fluctuations on transport in antiferromagnetic semimetals}

\author{Youngseok Kim}\UIUCECE\UIUCMNTL
\author{Moon Jip Park}\UIUCPHYS
\author{David G. Cahill}\UIUCMAT
 \author{Matthew J. Gilbert}\UIUCECE\UIUCMNTL\STANECE

\date{\today}

\begin{abstract}
Recent demonstrations on manipulating antiferromagnetic (AF) order have triggered a growing interest in antiferromagnetic metal (AFM), and potential high-density spintronic applications demand further improvements in the anisotropic magnetoresistance (AMR). The antiferromagnetic semimetals (AFS) are newly discovered materials that possess massless Dirac fermions that are protected by the crystalline symmetries. In this material, a reorientation of the AF order may break the underlying symmetries and induce a finite energy gap. As such, the possible phase transition from the semimetallic to insulating phase gives us a choice for a wide range of resistance ensuring a large AMR. To further understand the robustness of the phase transition, we study thermal fluctuations of the AF order in AFS at a finite temperature.
For macroscopic samples, we find that the thermal fluctuations effectively decrease the magnitude of the AF order by renormalizing the effective Hamiltonian. Our finding suggests that the insulating phase exhibits a gap narrowing at elevated temperatures, which leads to a substantial decrease in AMR.
We also examine spatially correlated thermal fluctuations for microscopic samples by solving the microscopic Landau-Lifshitz-Gilbert equation finding a qualitative difference of the gap narrowing in the insulating phase. For both cases, the semimetallic phase shows a minimal change in its transmission spectrum illustrating the robustness of the symmetry protected states in AFS. Our finding may serve as a guideline for estimating and maximizing AMR of the AFS samples at elevated temperatures.
\end{abstract}


\maketitle

\section{Introduction} \label{sec:intro}

The effort of understanding magnetic materials have triggered to discover a wide range of new physics which has been exploited to improve existing technologies\cite{Gibbs2004, Bruck2008, Gutfleisch2011, Dieny2017}. One of the prominent example is the solid-state magnetic memory\cite{Huai2008, Apalkov2016, Dieny2017} for its superior speed in writing operation\cite{Rasing2010, Jan2016}, the robustness on radiation-induced errors\cite{Sakimura2014}, and the high thermal stability\cite{Sato2012}. However, the magnetic moment in ferromagnetic materials is susceptible to the external magnetic field and often causes an undesirable bias toward one particular magnetic configuration due to the stray field\cite{Klostermann2002} from adjacent ferromagnetic films. 
An antiferromagnetic metals (AFM), having a zero net magnetic moment, are immune to the external magnetic field, thus providing an intrinsic advantage for realizing high-density spintronic devices\cite{Jungwirth2016}. 
In addition, a recent study on the current-induced spin-orbit torque\cite{Gambardella2011,Brataas2014} has established a promising way to manipulate antiferromagnetic (AF) orders by an electric current\cite{Jungwirth2014,Wadley2016}, and triggered a renewed interest in AFM\cite{Jungwirth2016,Tserkovnyak2018}. The anisotropic Fermi surface of AFM leads to changes in the current flow upon manipulation of the AF order. The subsequent difference between high and low resistance state determines the anisotropic magnetoresistance ratio (AMR), and has been served as a convenient read-out mechanism for AFM\cite{Wadley2016}. However, a typical AMR range found in AFM is a few percent\cite{Ramesh2014,Wadley2016} and may be overshadowed by random resistance variations in devices and external circuitaries for high-density integration\cite{Parkin2003}. Moreover, the resistance of the AFM is low in nature due to its metallic phase, and is unfavorable for an integration with conventional device technologies\cite{Apalkov2016}.

For this type of application, one particularly promising class of the material is the antiferromagnetic semimetals (AFS). 
In this study, we particularly focus on one type of AFS that possesses the Dirac fermion protected by an additional non-symmorphic crystalline symmetry. The Dirac fermion is characterized by the Dirac point\cite{Tang2016, Wang2017_1} or Dirac nodal line\cite{Wang2017_2} where the valence and conduction band touch.
In general, the protected crossings in Dirac semimetals require a presence of the inversion ($\mathcal{P}$) and the time-reversal ($\mathcal{T}$) symmetry. 
Although the $\mathcal{T}$ is absent in magnetic materials, the existence of the AF order enables the system to preserve the combined symmetry, $\mathcal{P}\mathcal{T}$, which leads to the discovery of new magnetic materials that the magnetism and the massless Dirac fermion coexist\cite{Tang2016}. As a result, the Dirac quasi-particles may be found in antiferromagnetic materials such as CuMnAs or CuMnP\cite{Tang2016}. 
The basic idea for the application is based on the observation that the AFS gains a finite energy gap when the underlying symmetry is broken by reorienting the AF order. Such transition from the semimetallic to insulating phase is dubbed as a topological (semi)metal-insulator transition (MIT)\cite{Jungwirth2017} and in principle provides a large ratio in magnetoresistance.

The discovery of the AFS calls for the further understanding of the behavior of the Dirac fermion in the presence of the magnetism. One intrinsic aspect of the magnetism that lacks the understanding in this particular material is the influence of the thermal fluctuation. 
We particularly aim to investigate the thermal fluctuation of the spin as (i) it may divert the antiferromagnetic order from its original orientation breaking the underlying non-symmorphic symmetry, and (ii) such thermal fluctuation always exists due to the inevitable coupling between a sample and its environment, especially at the temperature where realistic applications are considered. The goal of this study is to analyze the impact of random spin fluctuations in AFS. For this purpose, we first introduce a tight-binding Hamiltonian for AFS in Section~\ref{sec:model}. In Section~\ref{sec:configave}, we evaluate the impact of the thermal fluctuation in AF order orientation for a macroscopic sample by treating the fluctuation as an impurity potential. The subsequent changes in quasi-particle spectrum have been examined by averaging the impurity potential over all the possible configurations and we numerically calculate the resultant transport response. In Section~\ref{sec:LLG}, we consider a spatially correlated thermal fluctuation which has been neglected in macroscopic considerations. We consider the correlated fluctuations by solving the stochastic Landau-Lifshitz-Gilbert equation and the corresponding current response has been calculated using the non-equilibrium Green function formalism. The result has been compared with our previous results in Section~\ref{sec:configave} and both similarities and differences have been discussed. In Section~\ref{sec:conclusion}, we summarize our findings and conclude this study.

\section{Antiferromagnetic semimetal Hamiltonian} \label{sec:model}

 \begin{figure}[b!]
  \centering
   \includegraphics[width=0.5\textwidth]{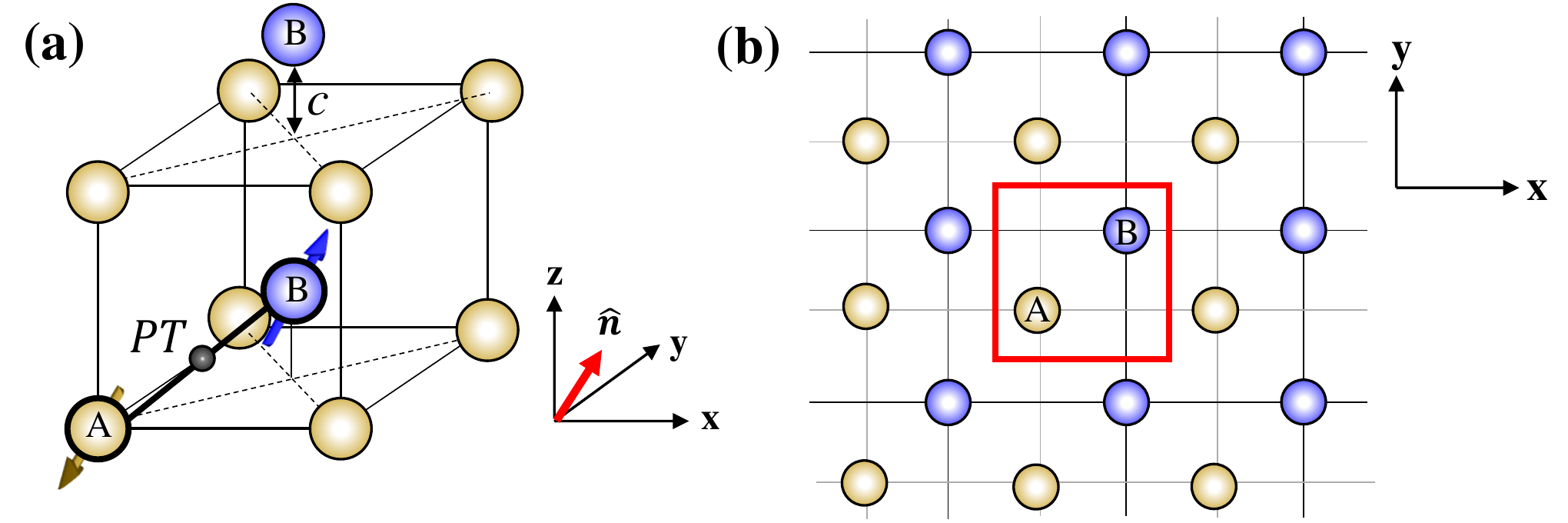}
  \caption{(a) A lattice structure consists of two sublattice atoms. Two sublattice atoms are indicated as A and B having an opposite spin configuration along the N\'{e}el vector $\hat{\bm{n}}$ (AF order) enforced by the exchange coupling. (b) The lattice structure has non-symmorphic symmetry $\mathcal{G}_x$ for $\hat{\bm{n}}||[100]$ and $\mathcal{G}_y$ for $\hat{\bm{n}}||[010]$.  }\label{fig:lattice}
\end{figure}
We begin our analysis by introducing a tight-binding model for AFS. Figure~\ref{fig:lattice}(a) shows a tetragonal primitive lattice structure\cite{Wang2017_2} in the space group 59 ($Pmmn$) consisting of the $A$ and $B$ sublattices. The spin degree of freedoms in $A$ and $B$ sublattices are anti-aligned and form an AF order, or specifically, a N\'{e}el order. The N\'{e}el order is characterized by a unit vector defined as the N\'eel vector $\hat{\bm{n}}$ and indicated as a red arrow in Fig.~\ref{fig:lattice}(a). The relative location of the $A$ and $B$ sublattice is determined by the inversion symmetry ($\mathcal{P}$) with respect to its inversion center located at the middle point of the adjacent $A$ and $B$ sublattices. Although $A$ and $B$ sublattices consist of the identical primitive unit cell structure with the same atomic compositions, $\mathcal{P}$ is explicitly broken due to the AF order. In addition, the time-reversal symmetry ($\mathcal{T}$) is broken due to the non-zero magnetic order. Nevertheless, the system preserves the combined symmetry\cite{Wang2017_1, Wang2017_2, Tang2016} ($\mathcal{PT}$), whose symmetry center is indicated as a black dot in Fig.~\ref{fig:lattice}(a). The presence of the $\mathcal{PT}$ symmetry guarantees the two-fold degeneracy in the whole Brillouin zone.

For more detailed analysis, we use the real-space tight-binding Hamiltonian adapted by \citeauthor{Wang2017_2}\cite{Wang2017_2}:
\begin{equation} \label{eq:Htot}
H=H_0+H_{int},
\end{equation}
where $H_0$ is the single-particle Hamiltonian that has the spin-orbit coupling, and $H_{int}$ is the Hamiltonian that describes the exchange interactions between itinerant electron spins and localized magnetic moments at $A$ and $B$ sublattices. We consider the single-particle Hamiltonian consists of an equal number of $A$ and $B$ sublattices with a total number of lattice sites $2N$. Then the single-particle Hamiltonian is\cite{Wang2017_2}
\begin{equation} \label{eq:H0}
\begin{split}
H_0=&\sum_{\langle i j\rangle, \langle\langle i j \rangle\rangle}t_{ij}c_i^\dagger c_j 
+\sum_{\langle\langle i j \rangle\rangle,\langle l \rangle} c_i^\dagger i\lambda_{ij}(\hat{\mathbf{d}}_{il}\times \hat{\mathbf{d}}_{lj})\cdot \bm{\sigma} c_j \\
\end{split}
\end{equation}
where $c_i^\dagger=(c_{i\uparrow}^\dagger,c_{i\downarrow}^\dagger)^T$ is the electron creation operator at site $i\in\{1,2,\cdots,2N\}$, $\langle ij \rangle$ stands for the nearest-neighbor site pair (adjacent $A$-$B$ sublattices) and $\langle\langle i j \rangle\rangle$ indicates the next-nearest-neighbor site pair (adjacent $A$-$A$ or $B$-$B$ sublattices) with a corresponding hopping parameter $t_{ij}$. The second term in Eq.~(\ref{eq:H0}) is the next-nearest hopping spin-orbit coupling (SOC)\cite{Kane2005, Jairo2017}. In Eq.~(\ref{eq:H0}), $\hat{\bm{d}}_{il}=(\mathbf{r}_i-\mathbf{r}_l)/\vert\mathbf{r}_i-\mathbf{r}_l\vert$ is a unit vector which connects the target atomic site $i$ (or $j$) with the intermediate atomic site $l$, which is positioned at the same separation from the site $i$ and $j$. $\bm{\sigma}=(\sigma_x,\sigma_y,\sigma_z)$ are the Pauli matrices for spin degree of freedom.
The interaction Hamiltonian is\cite{Maekawa2018}
\begin{equation} \label{eq:int}
H_{int}=\sum_{i} J\bm{{m}}_i \cdot \bm{s}_{i}+\sum_{\langle ij \rangle}A_{ex}\bm{{m}}_i\cdot \bm{{m}}_j,
\end{equation}
where the spin density operator is defined as $\bm{s}_i=c_i^\dagger \bm{\sigma} c_i$. Under the mean-field approximation within a unit cell, a local magnetization $\bm{{m}}_i$ is assumed at the site $i$ with a uniform saturation magnetization $\vert\bm{{m}}_i\vert=m_s$ for all lattice sites. In Eq.~(\ref{eq:int}), $J$ is the on-site exchange coupling constant between the itenerant electron spin ($\bm{s}_{i}$) and the local magnetic moment ($\bm{m}_{i}$), and $A_{ex}>0$ is the antiferromagnetic exchange constant between adjacent local magnetic moments. To further simplify the Hamiltonian, we define a unit cell structure consists of $A$ and $B$ sublattices, with a unit cell index $r\in\{1,2,\cdots,N\}$. The red rectangular box in Fig.~\ref{fig:lattice}(b) describes the $r$th unit cell whose localized spin angular momentum orientations are defined by $\bm{{m}}_r^A$ and $\bm{{m}}_r^B$ for $A$ and $B$ sublattices, respectively. We then assume that the antiferromagnetic exchange strength is strong enough to maintain an AF order within each unit cell by satisfying the condition $\bm{{m}}_r^A=-\bm{{m}}_r^B$. Here, we define $J\bm{{m}}_r^A=\Delta_r\bm{\hat{n}}_{r}$, where the N\'eel vector $\bm{\hat{n}}_{r}$ and the exchange energy $\Delta_r$ characterizes the orientation and the magnitude of the AF order, respectively. Consequently, Eq.~(\ref{eq:int}) becomes
\begin{equation} \label{eq:intAF}
H_{int}^{AF}=\sum_{r} \Delta_r\bm{\hat{n}}_{r}\cdot[ \mathbf{s}_r^A -\mathbf{s}_r^B],
\end{equation}
where $\bm{s}_r^{A/B}$ is a spin density operator of the A/B sublattice at the $r$th unit cell. The contribution of the local magnetic moments, or the second term in Eq.~(\ref{eq:int}), becomes constant for the assumed antiferromagnetic order merely adding a constant shift in the total Hamiltonian, therefore we may ignore its contribution in Eq.~(\ref{eq:intAF}). 
The more detailed discussions in Eq.~(\ref{eq:intAF}) is given in Appendix~\ref{app:Hint}.

In general, the mean-field approximation assumes that the local fluctuations of the magnetization are negligible, thus the local magnetic momentum is approximated by a global averaged value, or $J\bm{{m}}^A=(1/N)\sum_r J\bm{{m}}_r^A=\Delta \bm{\hat{n}}$, where we define the macroscopic N\'eel vector $\bm{\hat{n}}$, and the macroscopic exchange energy $\Delta$. Using the mean-field approximation, we obtain the following Hamiltonian in momentum space\cite{Wang2017_2}:
\begin{equation} \label{eq:H3D}
\begin{split}
\mat{H}_{\dvec{n}}(\vec{k})=&[t_{xy}\tau_1+t_z(\tau_1\cos k_z+\tau_2\sin k_z)]\cos\frac{k_x}{2}\cos\frac{k_y}{2}\\
&+t'_{xy}(\cos k_x + \cos k_y) + t'_z\cos k_z + \Delta\tau_3 \mathbf{\sigma}\cdot \dvec{n} \\
&+(\lambda-\lambda_z\cos k_z)\tau_3(\sigma_2\sin k_x -\sigma_1 \sin k_y),
\end{split}
\end{equation}
where $t_{xy}$ and $t_z$ are the nearest neighbor hopping constants, $t'_{xy}$ and $t'_z$ are the next nearest neighbor hopping constants, $\lambda$ and $\lambda_z$ are the spin-orbit coupling strength, and $\tau_{i=1,2,3}$ and $\sigma_{i=1,2,3}$ are the Pauli matrices for sublattice and spin degree of freedom, respectively. Note that the lattice constant is set to $a=1$ for simplicity.
The Hamiltonian in Eq.~(\ref{eq:H3D}) is known to possess\cite{Wang2017_2} $\mathcal{PT}=i\tau_1\sigma_2\mathcal{K}$ symmetry in addition to the gliding (non-symmorphic) mirror symmetry $\mathcal{G}_{x}=\{M_x|\frac{1}{2}00\}$ or $\mathcal{G}_y=\{M_y|0\frac{1}{2}0\}$ depending on the N\'{e}el vector orientation $\dvec{n}$. Here, the non-symmorphic symmetry operator $\mathcal{G}_x$ ($\mathcal{G}_y$) acts on the Hamiltonian by applying a mirror symmetry operator $M_x$ ($M_y$) followed by a half-lattice vector translation operator $T=\{\frac{1}{2}00\}$ ($T=\{0\frac{1}{2}0\}$), and is expressed in a matrix form as
\begin{equation}
\mathcal{G}_{x}=
\begin{pmatrix}
i\sigma_1 & 0 \\
0 & ie^{ik_x}\sigma_1 \\
\end{pmatrix} ,\;
\mathcal{G}_{y}=
\begin{pmatrix}
i\sigma_2 & 0 \\
0 & ie^{ik_y}\sigma_2 \\
\end{pmatrix}.
\end{equation}
In Eq.~(\ref{eq:H3D}), the N\'{e}el vector orientation is defined as
\begin{equation} \label{eq:n}
\dvec{n}=(\cos\varphi_{n}\sin\theta_{n},\sin\varphi_{n}\sin\theta_{n},\cos\theta_{n})^T,
\end{equation}
where $\varphi_n$ and $\theta_n$ is the in-plane and out-of-plane angle of the N\'{e}el vector, respectively.
When the N\'{e}el vector configuration is in $\dvec{n}||[100]$ (or $\varphi_n=0,\;\theta_n=\pi/2$), $\mathcal{G}_x$ is respected at $k_x=0,\pi$ and we may find the protected Dirac points\cite{Wang2017_1} or Dirac nodal lines\cite{Wang2017_2} at $k_x=\pi$. Similarly, when the N\'{e}el vector orientation is $\dvec{n}||[010]$, $\mathcal{G}_y$ is respected at $k_y=0,\pi$ and we may find the protected Dirac points (or Dirac nodal lines) at $k_y=\pi$. In both cases, the linear crossings at the edge of the Brillouin zone are protected by the underlying symmetries and the system may exhibit the gapless (semimetallic) phase. However, other N\'eel vector configurations (e.g. $\dvec{n}||[001]$) break the gliding mirror symmetries and the accidental band crossings are not protected. In this case, the system may show the gapped (insulating) phase.

The ground state configuration of the N\'eel vector may be obtained by evaluating the total free energy of possible N\'eel vector configurations. The result may vary as a function of the material parameters and the location of the chemical potential\cite{Kim2017}. 
As our goal is to understand the impact of the thermal fluctuation on the given AFS phase, we focus on a particular set of the assumed ground state N\'eel vector configurations which respect or break the underlying symmetry. The former case results in the gapless phase, whereas the latter scenario exhibits the gapped phase. In particular, we use $\dvec{n}||[100]$ (or $\varphi_n=0,\;\theta_n=\pi/2$) as a representative configuration for the gapless phase and $\mat{n}||[001]$ (or $\theta_n=0$) for the gapped phase. The detailed spectrum, low-energy Hamiltonian, and discussions on the parameter space of the model Hamiltonian may be found in \citeauthor{Wang2017_2}\cite{Wang2017_2} and \citeauthor{Kim2017}\cite{Kim2017}.

\section{Self-averaging N\'eel vector fluctuation}  \label{sec:configave}

\subsection{Thermal fluctuation of the N\'{e}el vector as an impurity potential}  \label{sec:Hxselfenergy}
The electrons in semimetals have ultrafast relaxation time scales due to the fact that (i) the Coulomb interaction is less screened in semimetals\cite{Brida2013,Mazhar2017}, and (ii) the linear dispersion ensures efficient electron-phonon scatterings\cite{Hofmann2013} including Auger type processes\cite{Winzer2010}. For example, typical semimetallic materials such as graphene\cite{Wolf2005,Winzer2010,Hofmann2013,Brida2013} and topological semimetals\cite{Mazhar2017} exhibit the electron relaxation time scale of $30$ - $500$ fs. This is a very short time scale compared to the antiferromagnetic precession time scale\cite{Ivanov2014, Tserkovnyak2018} in a range of few ps. For this reason, we assume that electronic bands of the AFS immediately respond to the dynamical changes of the magnetism. In other words, electrons see the macroscopic thermal fluctuation as a static object. This assumption allows us to treat the thermal fluctuations of the AFM as static disorders with a given magnetic configuration.

\begin{figure}[t!]
  \centering
   \includegraphics[width=0.5\textwidth]{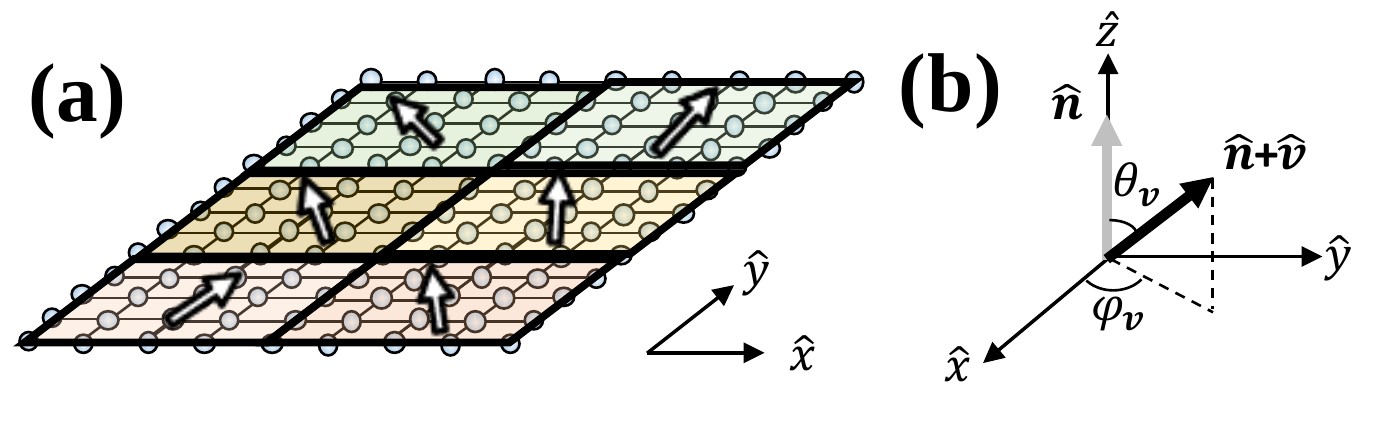}
  \caption{(a) A system comprised of a set of subsystems (patches) with an averaged N\'eel vector indicated as an arrow. Note that each patch contains enough unit cells so that the region is well described by momentum space Hamiltonian in Eq.~(\ref{eq:H3D}). 
(b) In the presence of a random thermal fluctuation, we assume that the N\'{e}el vector orientation deviates from the original configuration by an azimuthal angle $\varphi_v$ and polar angle $\theta_v$. The original N\'eel vector orientation is aligned with $\hat{z}$ direction as a gray solid arrow, and a possible N\'{e}el vector deviation is described by a black solid arrow.
  }\label{fig:schematics}
\end{figure}

Especially at a sufficiently high temperature, the coherence length of the electron is much smaller than the sample size. In this case, we may think of the system consisting of numerous phase-independent sub-systems. The observables of such system are effectively evaluated by averaging over particular impurity configurations at each independent sub-system\cite{Bruus2004}, and such procedure is often referred to as \emph{self-averaging}. The self-averaging over possible impurity configurations has been used to obtain the renormalized Hamiltonian in the presence of random on-site impurities\cite{Altland2010,Park2017}. Figure~\ref{fig:schematics}(a) illustrates the situation where the system comprised of patches of sub-systems (rectangular solid lines) with the corresponding macroscopic N\'eel vector fluctuations. The self-averaging procedure on such a system provides an analytical form of the renormalized Hamiltonian and an intuitive way of understanding the impact of the N\'eel vector fluctuation. Therefore, we exploit the self-averaging method in this section to discuss the possible consequences of the N\'eel vector fluctuations for the gapped and gapless phase of AFS.

We begin with evaluating the N\'eel vector fluctuation by self-averaging the sub-systems that have different N\'eel vector fluctuations. One way to obtain the changes in the spectrum and observables is to consider the Green function of the unperturbed system and take account for the fluctuations perturbatively. In this regard, the fluctuation in the N\'eel vector is considered as an impurity potential.
Figure~\ref{fig:schematics}(b) shows the N\'{e}el vector deviation from its spatially averaged value, $\bm{\hat{n}}$.
To describe the deviation of the N\'{e}el vector, we define a following perturbation on the N\'{e}el vector at $v$th sub-system:
\begin{equation} \label{eq:v}
\dvec{v}(\varphi_v,\theta_v)
=(\cos\varphi_v\sin\theta_v,\sin\varphi_v\sin\theta_v,\cos\theta_v-1)^T,
\end{equation}
where $\theta_v$ and $\varphi_v$ are a polar and azimuthal angle with respect to the $\hat{z}$ axis, respectively. The fluctuation vector given in Eq.~(\ref{eq:v}) takes account for a deviation from the averaged N\'{e}el vector that is assumed to be aligned in $\hat{z}$ direction. We may incorporate Eqs.~(\ref{eq:n}) and (\ref{eq:v}) by properly rotating the $\hat{z}$ axis of $\dvec{v}$ to align with $\dvec{n}$. Then, we define the fluctuation Hamiltonian
\begin{equation} \label{eq:RRv}
\mat{V}=\Delta\tau_3\bm{\sigma}\cdot R_z(\varphi_n)R_y(\theta_n)\dvec{v},
\end{equation}
where $R_z$ and $R_y$ are the rotational operator about the $\hat{z}$ and $\hat{y}$ axis, respectively. As a result, the Hamiltonian in Eq.~(\ref{eq:H3D}) is modified as $\mat{H}_\dvec{n}\rightarrow \mat{H}_\dvec{n}+\mat{V}$ and
$
\Delta\tau_3\bm{\sigma}\cdot\dvec{n}\rightarrow
\Delta\tau_3\bm{\sigma}\cdot  [\dvec{n}+R_z(\varphi_n)R_y(\theta_n)\dvec{v}]
$,
which takes account for a deviation of the N\'{e}el vector by treating $\mat{V}$ as an impurity potential. 
As a result, thermal fluctuations of the macroscopic magnets redistribute the Neel vector. Assuming an uniaxial anisotropy energy along the averaged N\'eel vector orientation (e.g. $\hat{z}$ axis), the probability density function of the fluctuation polar and azimuthal angle along the anisotropy axis follows\cite{Butler2012}
\begin{equation} \label{eq:Dtheta}
D_{\theta_v} \propto \exp(-U_{th}\sin^2\theta_v),\; D_{\varphi_v}=\frac{1}{2\pi},
\end{equation}
respectively, where the probability density functions satisfy $\int_0^{\pi}\sin\theta_vd\theta_v \int_0^{2\pi}d\varphi_v D_{\theta_v}D_{\varphi_v}=1$. In Eq.~(\ref{eq:Dtheta}), the normalized thermal barrier, $U_{th}$, is defined as\cite{Chshiev2017}
\begin{equation} \label{eq:Uth}
U_{th}=\frac{K_u V}{k_B T}
\end{equation}
where $K_u$ is the anisotropy energy of the macroscopic magnet, $V$ is the volume of the grain, $k_B$ is Boltzmann constant, and $T$ is the temperature.

We now wish to consider the fluctuation Hamiltonian $\mat{V}$ perturbatively and evaluate its impact on the Dirac quasi-particles. According to the effective medium theory\cite{Sheng1995}, we obtain the self-energy of the impurity potential up to the second order in $\mat{V}$ as
\begin{equation} \label{eq:Sigeff_approx}
\begin{split}
\mat{\Sigma}_{eff}
=&\int \sin\theta_v d\theta_v d\varphi D_{\theta_v} D_{\varphi_v} \mat{V}(\mat{I}-\mat{G}_{eff}[\mat{V}-\mat{\Sigma}_{eff}])^{-1} \\
\simeq&
\int \sin\theta_v d\theta_v d\varphi_v D_{\theta_v} D_{\varphi_v} \mat{V}+\mat{V}\mat{G}_{eff}[\mat{V}-\mat\Sigma_{eff}],
\end{split}
\end{equation}
where $\mat{G}_{eff}=[\epsilon-\mat{H}_0-\mat{\Sigma}_{eff}]^{-1}$ is the effective Green function of the system at a given energy $\epsilon$. In Eq.~(\ref{eq:Sigeff_approx}), the self-averaging procedure is carried out by considering all the possible fluctuation configurations with the corresponding probability density functions $D_{\theta_v}$ and $D_{\varphi_v}$, and obtain the self-averaged self-energy $\mat{\Sigma}_{eff}$. The detailed derivation of Eq.~(\ref{eq:Sigeff_approx}) is given in Appendix~\ref{sec:effective_medium}. To obtain a closed form of $\mat{\Sigma}_{eff}$, we approximate the effective Green function as a bare Green function, or $\mat{G}_{eff}\simeq \mat{G}_0=[\epsilon-\mat{H}_0]^{-1}$ and ignore $\mat\Sigma_{eff}$ in the right hand side of Eq.~(\ref{eq:Sigeff_approx}). As a result, we obtain the self-energy in the second-order Born approximation form:
\begin{equation} \label{eq:Sigeff2}
\begin{split}
\mat\Sigma_{eff}\simeq&
\int \sin\theta_v d\theta_v d\varphi D_{\theta_v} D_{\varphi_v} [\mat{V}+\mat{V}\mat{G}_{0}\mat{V}] \\
=&\braket{\mat{V}}+\braket{\mat{V}\mat{G}_{0}\mat{V}} \\
=&\Sigma_0+\Sigma_1\tau_1+\Sigma_2\tau_2
+\Sigma_3\tau_3\sigma_1
+\Sigma_4\tau_3\sigma_2
+\Sigma_5\tau_3\sigma_3,
\end{split}
\end{equation}
where $\braket{\cdot}$ stands for an average over all the possible fluctuation configurations.
In the last equality of Eq.~(\ref{eq:Sigeff2}), $\mat{\Sigma}_{eff}$ is expressed in terms of the all possible gamma matrices with the corresponding self-energy terms $\Sigma_{0,1,2,3,4,5}$. The obtained self-energy terms renormalize the parameters in the Hamiltonian and may give us insight into how thermal fluctuation changes the quasi-particle spectrum.

\subsection{Bare Green function, $\mat{G}_0$}
We now aim to evaluate the self-energy of the fluctuation Hamiltonian in Eq.~(\ref{eq:Sigeff2}). To this end, we first need to evaluate a bare Green function $\mat{G}_0$ in terms of relevant parameters in Eq.~(\ref{eq:H3D}). For notational simplicity, we rewrite the Hamiltonian in Eq.~(\ref{eq:H3D}) as
\begin{equation} \label{eq:Hsimple}
\mat{H}_\dvec{n}(\vec{k})=a_0+a_1\tau_1+a_2\tau_2+a_3\tau_3\sigma_1+a_4\tau_3\sigma_2+a_5\tau_3\sigma_3,
\end{equation}
where
\begin{equation} \label{eq:a012345}
\begin{split}
a_0=&t'_{xy}(\cos k_x + \cos k_y) + t'_z\cos k_z , \\
a_1=&(t_{xy}+t_z\cos k_z)\cos\frac{k_x}{2}\cos\frac{k_y}{2}, \\
a_2=&t_z\sin k_z\cos\frac{k_x}{2}\cos\frac{k_y}{2}, \\
a_3=&-(\lambda-\lambda_z\cos k_z)\sin k_y+\Delta\cos\varphi_n\sin\theta_n, \\
a_4=&(\lambda-\lambda_z\cos k_z)\sin k_x+\Delta\sin\varphi_n\sin\theta_n, \\
a_5=&\Delta \cos\theta_{n}. \\
\end{split}
\end{equation}
The retarded Green function for the bare Hamiltonian $\mat{H}_0$ is defined as
\begin{equation} \label{eq:G0}
 \begin{split}
 \mat{G}_0=&(\epsilon-\mat{H}_0+i0^+)^{-1}= \mat{G}_{0-}+\mat{G}_{0+},
 \end{split}
 \end{equation}
where $0^+$ is an infinitesimal positive number and $\epsilon$ is an energy. In Eq.~(\ref{eq:G0}),
 \begin{equation}  \label{eq:Gpm}
 \begin{split}
 &\mat{G}_{0\pm}=\frac{1}{2}\frac{1}{\epsilon-(a_0\pm a)+i0^+}\times \\
 &\left(
 1\pm \frac{1}{a}
 (a_1\tau_1
+a_2\tau_2
+ a_3\tau_3\sigma_1
+a_4\tau_3\sigma_2
+a_5\tau_3\sigma_3)
 \right), \\
 \end{split}
 \end{equation}
 where $a=\sqrt{a_1^2+a_2^2+a_3^2+a_4^2+a_5^2}$. The algebraic procedures to obtain Eq.~(\ref{eq:G0}) is summarized in Appendix~\ref{sec:Ham_and_eig}. With our knowledge of the bare Green function, we compute the self-energy for a gapless ($\dvec{n}||[100]$) and gapped ($\dvec{n}||[001]$) phase and evaluate the impact of the thermal fluctuation on their spectra.

\subsection{N\'{e}el vector fluctuation in gapless phases}\label{sec:renormalize_gapless}

Assuming that the N\'{e}el vector is initially aligned with $[100]$ direction, the Hamiltonian in Eq.~(\ref{eq:H3D}) is reduced to
\begin{equation} \label{eq:Hx}
\begin{split}
\mat{H}_{\dvec{n}||[100]}(\vec{k})=&a_0+a_1\tau_1+a_2\tau_2+a_3\tau_3\sigma_1+a_4\tau_3\sigma_2,
\end{split}
\end{equation}
where $a_{0,1,2}$ are the same as in Eq.~(\ref{eq:a012345}), but now we have $a_3=-(\lambda-\lambda_z\cos k_z)\sin k_y + \Delta$, $a_4=(\lambda-\lambda_z\cos k_z)\sin k_x$, and $a_5=0$. The Hamiltonian in Eq.~(\ref{eq:Hx}) respects $\mathcal{PT}$ symmetry as well as $\mathcal{G}_x$, which protects the linear crossings of the bands at the edge of the Brillouin zone\cite{Wang2017_1,Wang2017_2}, or at $k_x=\pi$.

Following our previous discussions, we obtain the Green function by taking account for the parameters in Eq.~(\ref{eq:Hx}), and the impurity potential $\mat{V}$ in Eq.~(\ref{eq:RRv}) for the specific N\'{e}el vector under consideration by using $\theta_n=\pi/2$ and $\varphi_n=0$. Then, we obtain the self-energy in Eq.~(\ref{eq:Sigeff2}). The detailed algebraic procedures for the self-energy calculation is outlined in Appendix~\ref{sec:evaluation} and \ref{sec:VxyzCalc}. 
The imaginary part of the obtained self-energy broadens the eigenvalue spectrum, whereas the real part effectively renormalizes the dispersion of the Hamiltonian. As our main interest is to see the shift of the eigenvalue spectrum and subsequent changes in gapped or gapless spectrum of the Hamiltonian, we only focus on the real part of the self-energy for below analysis.
Including the real part of the self-energy, the Hamiltonian in Eq.~(\ref{eq:Hx}) is renormalized as follows:
\begin{equation} \label{eq:Hgapless}
\begin{split}
& \mat{H}_{\dvec{n}||[100]}
=a_0+a_1\tau_1+a_2\tau_2+a_3\tau_3\sigma_1+a_4\tau_3\sigma_2 \\
\rightarrow& \mat{H}_{\dvec{n}||[100]}+\text{Re}\{\mat\Sigma\} \\
& \quad\quad\quad\;\;
=a'_0+a'_1\tau_1+a_2\tau_2+a'_3\tau_3\sigma_1+a_4\tau_3\sigma_2, \\
\end{split}
\end{equation}
where
\begin{equation} \label{eq:a_gapless}
\begin{split}
a'_{0,1,3}=&a_{0,1,3}+\text{Re}\{\Sigma_{0,1,3}\}.
\end{split}
\end{equation}
Each component of the self-energy is
\begin{equation} \label{eq:Sigma_gapless}
\begin{split}
\Sigma_0=& (A_{0+}+A_{0-})\Delta^2 (f_x+f_{yz}), \\
\Sigma_{1}=& -(A_{1+}-A_{1-})\Delta^2 (f_x+f_{yz}), \\
\Sigma_3=&  (A_{3+}-A_{3-})\Delta^2 (f_x-f_{yz}) -\Delta f_{x0} , \\
\end{split}
\end{equation}
where $f_x=\int_{0}^{\pi} d\theta D_\theta \sin\theta (1-\cos\theta)^2 $, $f_{yz}=\int_{0}^{\pi} d\theta D_\theta \sin^3\theta$, $f_{x0}=\int_{0}^{\pi} d\theta D_\theta \sin\theta(1-\cos\theta)$, and the numerical coefficients are
\begin{equation} \label{eq:A}
\begin{split}
A_{0\pm}=&\int_{BZ} \frac{d^3k}{(2\pi)^3}
 \frac{1}{\epsilon-(a_0\pm a)+i0^+}\frac{1}{2}, \\
A_{i\pm}=& \int_{BZ} \frac{d^3k}{(2\pi)^3}
 \frac{1}{\epsilon-(a_0\pm a)+i0^+}\frac{1}{2}\frac{a_i}{a}, \\
\end{split}
\end{equation}
for $i\in\{1,2,3,4,5\}$.
Note that $\Sigma_0$ merely renormalizes an on-site term and may induce a constant shift of the overall spectrum. As we are interested in the stability of the MIT, we focus on $\Sigma_{1,2,3,4,5}$ which renormalizes the dispersion of the system. Further analysis shows that $\Sigma_{2,4,5}$ terms vanish for this particular Hamiltonian up to the second order correction. See Appendix~\ref{sec:VxyzCalc} for the detailed calculation of the self-energy terms. 

In order to understand the relative contribution of $\Sigma_1$ to the low-energy spectrum, we may consider an additional mass term $m_1\tau_1$ induced by $\Sigma_1$ in the presence of the protected linear crossings at the edge of the Brillouin zone. Such mass term does not commute with the glide mirror symmetry at $k_x=\pi$, or $G_x=i\tau_3\sigma_1$, thus breaks the underlying symmetry and opens a gap in the quasi-particle spectrum. Meanwhile, the mass term $m_3\tau_3\sigma_1$ acquired from $\Sigma_3$ merely renormalizes the exchange energy as $\Delta\rightarrow\Delta'=\Delta+m_3$ and shifts the location of the linear crossings in the momentum space maintaining the gapless spectrum. Equation~(\ref{eq:Sigma_gapless}) shows that the $\Sigma_3$ term renormalizes the $a_3$ term in the first order of $\Delta$, whereas the lowest order correction of the other terms are in the second order of $\Delta$. Thus, we find that $\Sigma_3$ term plays a major role in renormalizing the Hamiltonian for small $\Delta$ limit, or $\Delta\ll 1$. Similar argument holds when the thermal barrier $U_{th}$ in Eq.~(\ref{eq:Uth}) is sufficiently large. When $U_{th}\gg 1$, the exponential probability distribution $D_\theta$ in Eq.~(\ref{eq:Dtheta}) only allows a small N\'eel vector fluctuation in the polar angle, or $\theta\sim 0$. By linearizing $\sin\theta\simeq \theta$ and $1-\cos\theta\simeq \theta^2/2$ for $f_{x,yz,x0}$ in Eq.~(\ref{eq:Sigma_gapless}), higher order corrections on $\mat{V}$ possesses higher order terms in $\theta$. Therefore, the lowest correction term plays a significant role in renormalizing the Hamiltonian.

\begin{figure}[t!]
  \centering
   \includegraphics[width=0.5\textwidth]{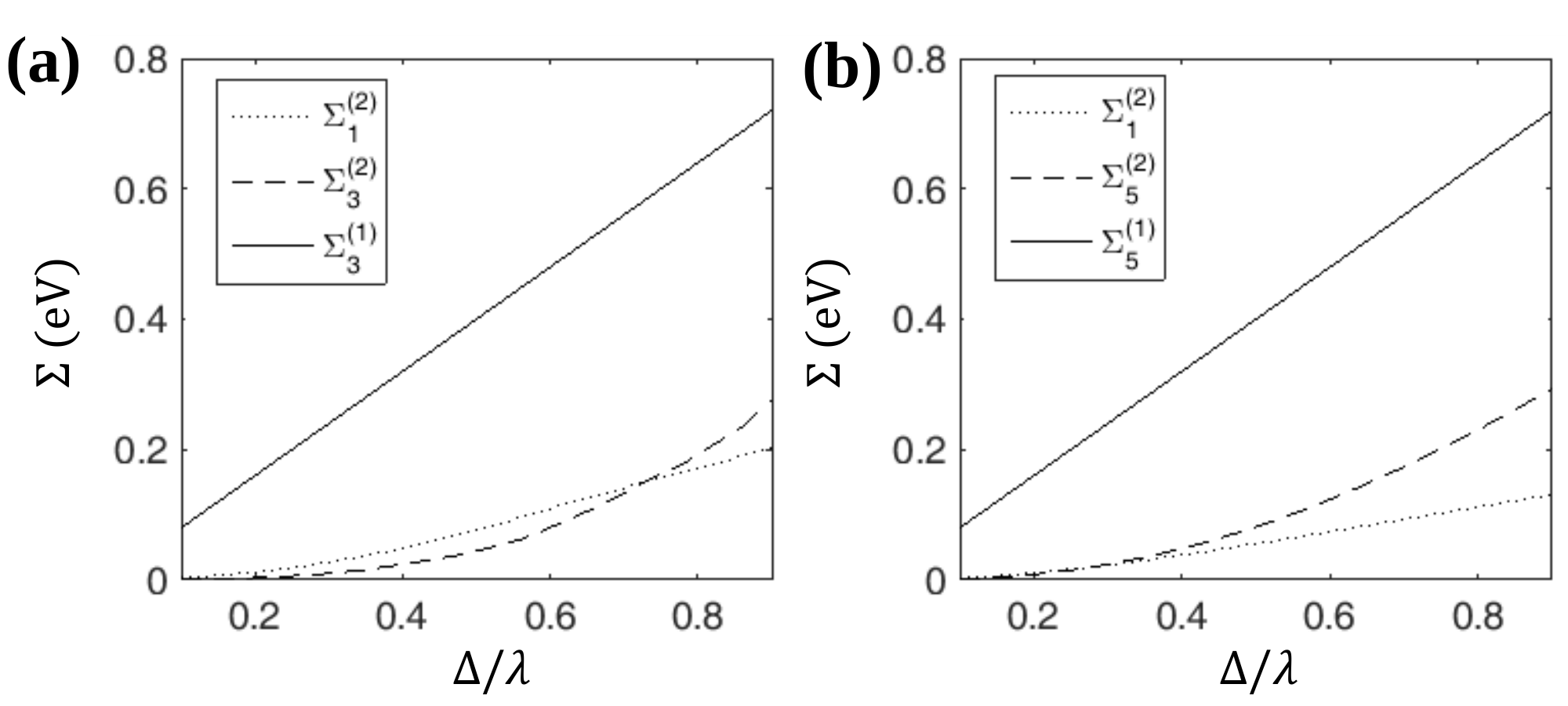}
  \caption{(a) The numerically computed coefficients of the impurity potential self-energy when the averaged N\'eel vector is $\dvec{n}||[100]$ (gapless). The mass term introduced by $\Sigma_1$ may gap out the Hamiltonian, whereas $\Sigma_3$ merely shifts the Dirac points in momentum space. The dotted line shows $\Sigma_1^{(2)}=(A_{1+}-A_{1-})\Delta^2$, which is the lowest correction term for $\Sigma_1$. The dashed line and solid line show $\Sigma_3^{(2)}=(A_{3+}-A_{3-})\Delta^2$ and $\Sigma_3^{(1)}=\Delta$ which corresponds to the second order and first order correction term in $\Sigma_3$, respectively.
(b) The same calculation, but the averaged N\'eel vector is $\dvec{n}||[001]$ (gapped). The dotted line shows $\Sigma_1^{(2)}=(A_{1+}-A_{1-})\Delta^2$, which is the lowest correction term for $\Sigma_1$. The dashed line and solid line show $\Sigma_5^{(2)}=(A_{5+}-A_{5-})\Delta^2$ and $\Sigma_5^{(1)}=\Delta$ which corresponds to the second order and first order correction term in $\Sigma_5$, respectively.
The parameters used for simulation are $t=1$, $t_{xy}=t$, $t_{z}=0.5t$, $t'_{xy}=0.05t$, $t'_{z}=0.05t$, $\lambda=0.8t$, $\lambda_z=0.3t$, and $\Delta=0.3t$, which is identical with that used in \citeauthor{Wang2017_2}\cite{Wang2017_2}. We plot the results within the range $0<\Delta/\lambda<0.9$, where the system is guaranteed to have at least two Dirac nodal lines\cite{Wang2017_2}. 
  }\label{fig:Sigma}
\end{figure}

To obtain a quantitative understanding on relative magnitudes of each terms in Eq.~(\ref{eq:Sigma_gapless}), we evaluate the second order correction of $\Sigma_1$, the first order and the second order correction of $\Sigma_3$.
Fig.~\ref{fig:Sigma}(a) shows the calculated numerical coefficients in Eq.~(\ref{eq:Sigma_gapless}) by assuming $\epsilon=0$.
The second order correction of $\Sigma_1$, or $\Sigma_1^{(2)}=(A_{1+}-A_{1-})\Delta^2$, is plotted in a dotted line. The first order correction of $\Sigma_3$, or $\Sigma_3^{(1)}=\Delta$, and the second order correction of $\Sigma_3$, or $\Sigma_3^{(2)}=(A_{3+}-A_{3-})\Delta^2$, are plotted in a solid and dashed line, respectively.
As the system facilitates symmetry protected Dirac nodal lines\cite{Wang2017_2,Kim2017} for the specific parameter range of $|\Delta|<\lambda+\lambda_z$, we plot the results within the range of $0<\Delta<0.9\lambda$, where the system is guaranteed to have at least two Dirac nodal lines.
When $\Delta$ is small, the self-energy term is dominantly determined by $\Sigma_3^{(1)}$ and the second order corrections are negligible, thereby we may ignore $\Sigma_1$. In this case, $a'_3$ merely renormalizes the exchange energy and the spectrum remains gapless even in the presence of thermal fluctuations. It is important to note that this semimetallic picture is only valid when the system size is in the larger length scale than the size of the each fluctuating domain. Otherwise, we may observe a transport response for the finite gap obtained in each magnetic domain due to the fluctuation. In addition, even when the gapped domains are averaged out to exhibit gapless spectrum, it introduces a backscattering process which may appear as a finite life-time of quasi-particles.

\subsection{N\'{e}el vector fluctuation in gapped phases}\label{sec:renormalize_gapped}

We now examine the impact of the N\'{e}el vector fluctuation when the system is initially gapped. For this purpose, we assume $\hat{\mathbf{n}}||[001]$ and rewrite our system Hamiltonian as
\begin{equation} \label{eq:Hz}
 \mat{H}_{\dvec{n}||[001]}
 =a_0+a_1\tau_1+a_2\tau_2+a_3\tau_3\sigma_1+a_4\tau_3\sigma_2+a_5\tau_3\sigma_3,
\end{equation}
where $a_{0,1,2}$ are the same as in Eq.~(\ref{eq:a012345}), but now we have $a_3=-(\lambda-\lambda_z\cos k_z)\sin k_y$, $a_4=(\lambda-\lambda_z\cos k_z)\sin k_x$, and $a_5=\Delta$. Following similar procedures outlined in Section~\ref{sec:renormalize_gapless}, we set the Green function and impurity potential $\mat{V}$ in Eq.~(\ref{eq:RRv}) for the specific N\'{e}el vector under consideration by using $\theta_n=0$.
Following the procedure outlined in Section~\ref{sec:Hxselfenergy} and Appendix~\ref{sec:evaluation}-\ref{sec:VxyzCalc}, we find that $\Sigma_0$, $\Sigma_1$, and $\Sigma_5$ shows non-zero 
self-energies. We have vanishing $\Sigma_2$, $\Sigma_3$, and $\Sigma_4$ due to the fact that $a$ and $a_0$ are even function in momentum, but $a_2$, $a_3$, and $a_4$ are odd function in $k_z$, $k_y$, and $k_x$, respectively. As a result, the Hamiltonian is renormalized as follows:
\begin{equation}\label{eq:Hgapped}
\begin{split}
&  \mat{H}_{\dvec{n}||[001]}=a_0+a_1\tau_1+a_2\tau_2+a_3\tau_3\sigma_1
+a_4\tau_3\sigma_2+a_5\tau_3\sigma_3 \\
\rightarrow&  \mat{H}_{\dvec{n}||[001]}+\text{Re}\{\mat\Sigma\} \\
& \quad\quad\quad\;\;
=a'_0+a'_1\tau_1+a_2\tau_2+a_3\tau_3\sigma_1
+a_4\tau_3\sigma_2+a'_5\tau_3\sigma_3, \\
\end{split}
\end{equation}
where
\begin{equation} \label{eq:a015}
\begin{split}
a'_{0,1,5}=&a_{0,1,5}+\text{Re}\{\Sigma_{0,1,5}\}.
\end{split}
\end{equation}
Each component of the self-energy is
\begin{equation} \label{eq:Sigma_gapped}
\begin{split}
\Sigma_0=& (A_{0+}+A_{0-})\Delta^2 (f_z+f_{xy}), \\
\Sigma_1=& -(A_{1+}-A_{1-})\Delta^2 (f_z+f_{xy}), \\
\Sigma_5=&  (A_{5+}-A_{5-})\Delta^2 (f_z-f_{xy}) -\Delta f_{z0} , \\
\end{split}
\end{equation}
where $f_z=\int_{0}^{\pi} d\theta D_\theta \sin\theta(1-\cos\theta)^2 $, $f_{xy}=\int_{0}^{\pi} d\theta D_\theta \sin^3\theta$, and $f_{z0}=\int_{0}^{\pi} d\theta D_\theta \sin\theta(1-\cos\theta)$. Following similar arguments in Section~\ref{sec:renormalize_gapless}, we evaluate the coefficients of the self-energy in Eq.~(\ref{eq:Sigma_gapped}) to present its relative magnitudes of the first and second order corrections in $\Sigma_1$ and $\Sigma_5$. Fig.~\ref{fig:Sigma}(b) shows the second order correction of $\Sigma_1$, or $\Sigma_1^{(2)}=(A_{1+}-A_{1-})\Delta^2$, in a dotted line. Similarly, Fig.~\ref{fig:Sigma}(b) presents the first order correction of $\Sigma_5$, or $\Sigma_5^{(1)}=\Delta$, and the second order correction of $\Sigma_5$, or $\Sigma_5^{(2)}=(A_{3+}-A_{3-})\Delta^2$, in a solid and dashed line, respectively. The numerical result shows that the correction is dominantly determined by the first order term in $\Sigma_5$. As the gap size of the system is determined\cite{Wang2017_1, Kim2017} by $a_5$, the first order correction term in $a'_5$ directly alters the gap size of the system in the presence of the N\'{e}el vector fluctuation.

The gap size as a function of the N\'eel vector fluctuation may be evaluated by considering the thermal fluctuation angles $\theta_v$ and its exponential probability density function $D_{\theta}$ in Eq.~(\ref{eq:Dtheta}). A specific equation for the renormalized energy gap $a'_5$ may be obtained by performing the integration in $f_{z0}$. However, it is not straightforward to obtain an analytical expression for $a'_5$. Rather than numerically evaluate $a'_5$, we assume a simpler probability distribution function $D_{\theta}$ for an illustrative purpose.
We use a phenomenological parameter $\theta_{max}$, which sets the maximally allowed thermal fluctuation polar angle that characterizes the magnitude of the thermal fluctuations. We then set the $\theta_v$ to follow a uniform distribution over $\theta_v\in[0,\theta_{max}]$. Then, $a'_5$ up to the first order correction becomes
\begin{equation}\label{eq:a5}
a'_5\simeq \Delta\cos^2\frac{\theta_{max}}{2},
\end{equation}
where $0\leq\theta_{max}\leq \pi$. Equation~(\ref{eq:a5}) shows that the gap size is a decreasing function for an increasing thermal fluctuation magnitude. Therefore, we may observe an effective gap narrowing for increasing thermal fluctuations, when the averaged N\'{e}el vector orientation makes the system initially in the gapped phase. 

\subsection{Numerical results on spatially uncorrelated thermal fluctuations} \label{sec:transport_result}

 \begin{figure*}
  \centering
   \includegraphics[width=1.0\textwidth]{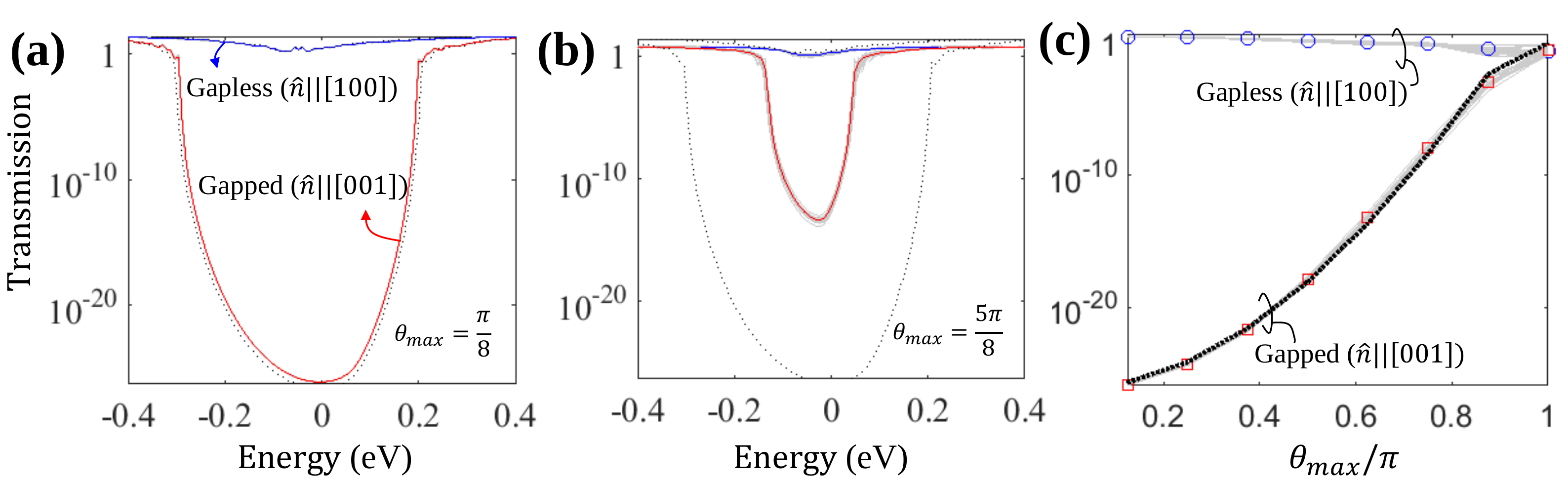}
  \caption{(a) A plot of transmission as a function of energy with the maximum fluctuation angle of $\theta_{max}=\pi/8$. 20 different random configurations are averaged and the resultant transmission is presented in red and blue line for gapped ($\hat{n}||[001]$) and gapless ($\hat{n}||[100]$) phase, respectively. The corresponding transmission without N\'{e}el vector fluctuation is plotted in dotted line for comparison. (b) The same plot with (a), but with the maximum angle of $\theta_{max}=5\pi/8$. 20 different random configurations are plotted in gray lines.
(c) A plot of transmission as a function of maximum N\'{e}el vector fluctuation angle $\theta_{max}$ at $E=-40$ meV. The gray lines indicate the results from 20 different random configurations and blue circular and red rectangular symbol are obtained by averaging those different configurations for gapless and gapped phase, respectively. The dotted line is a fitting results by using $T=f_Be^{-f_A\; \cos^2(\theta_{max}/2)}$, where $f_A$ and $f_B$ is a fitting parameter.
The parameters used for 2D Hamiltonian in Eq.~(\ref{eq:H2D}) are $t=1$, $t_{xy}=t$, $t'_{xy}=0.05t$, $\lambda=0.8t$, and $\Delta=0.3t$. The device dimension is fixed as $N_x\times N_y=50\times 100$, where $N_x$ is width and $N_y$ is length of the device.
  }\label{fig:transport}
\end{figure*}   

To examine our discussions in Sections~\ref{sec:renormalize_gapless} and \ref{sec:renormalize_gapped}, we construct a real-space Hamiltonian and numerically compute the transmission in the presence of the N\'{e}el vector fluctuation using the non-equilibrium Green function (NEGF) formalism\cite{Datta2005}. In particular, our numerical analysis follows the similar method that has been utilized for evaluating the impact of the charged impurities in topological insulator\cite{Rotter2013} or Weyl semimetals\cite{Xie2015,Park2017}.
In this work, we simplify our system Hamiltonian to be two-dimensional by setting $t_z=t'_z=\lambda_{z}=0$ to obtain 2D antiferromagnetic semimetal Hamiltonian\cite{Wang2017_2}
\begin{equation} \label{eq:H2D}
\begin{split}
\mat{H}_{\dvec{n}}^{2D}(\vec{k})=&t_{xy}\tau_1\cos\frac{k_x}{2}\cos\frac{k_y}{2}
+t'_{xy}(\cos k_x + \cos k_y) \\
&+ \Delta \tau_3 \mathbf{\sigma}\cdot \dvec{n} +\lambda\tau_3(\sigma_2\sin k_x -\sigma_1 \sin k_y).
\end{split}
\end{equation}
Nevertheless, the analysis shown in Sections~\ref{sec:renormalize_gapless} and \ref{sec:renormalize_gapped} are generally applicable both to 2D and 3D antiferromagnetic semimetals\cite{Wang2017_1, Wang2017_2}, as the Hamiltonian both in Eqs.~(\ref{eq:H3D}) and (\ref{eq:H2D}) preserves $\mathcal{PT}$ symmetry and non-symmorphic symmetries $\mathcal{G}_{x(y)}$ for particular N\'{e}el vector configurations. We then Fourier transform Hamiltonian in Eq.~(\ref{eq:H2D}) into the real-space, $\mat{H}_{\dvec{n}}^{2D}(\mathbf{r})$. 
We model the thermal fluctuation as a spatially uncorrelated N\'{e}el vector fluctuation, $\mat{V}(\mathbf{r})$, given in Eq.~(\ref{eq:RRv}) whose fluctuation azimuthal and polar angle are assumed to follow a uniform random distribution over the range of $\varphi_v\in(0,2\pi]$ and $\theta_v\in[0,\theta_{max}]$, respectively, at each site. Here, we utilize the phenomenological parameter $\theta_{max}$ to obtain a simplified form of the renormalized parameter $a'_5$ in Eq.~(\ref{eq:a5}). The analytical form of $a'_5$ provides an additional insight on our numerical analysis. In particular, we will show that the real part of the first order correction sufficiently captures the impact of the thermal fluctuation on the two-terminal transmission spectrum for an increasing thermal fluctuation magnitude. 

Having the total Hamiltonian, we construct the retarded Green function as
\begin{equation} \label{eq:Gr2D}
\mat{G}^r(E,\mathbf{r})=[(E+i\eta)-\mat{H}_{\dvec{n}}^{2D}(\mathbf{r})-\mat{V}(\mathbf{r})-\mat\Sigma_L-\mat\Sigma_R]^{-1},
\end{equation}
where $E$ is the energy and the infinitesimal broadening is set to $\eta=0.1$ meV. In Eq.~(\ref{eq:Gr2D}), we use wide-band limit approximation\cite{Datta2005} for left and right contact self-energies, which are defined as $\Sigma_{L}=-it_c\sum_{i=1}^{N_x} c^\dagger_{i1}c_{i1}$ and $\Sigma_{R}=-it_c\sum_{i=1}^{N_x} c^\dagger_{iN_y}c_{iN_y}$, respectively, where $c_{ij}$ ($c^\dagger_{ij}$) is an electron annihilation (creation) operator at site $\mathbf{r}=(i,j)$. The constant hopping parameter is set to $t_c=0.5$ eV, but the results are not dependent on a particular choice of $t_c$ as long as it is sufficiently large to inject the current into the sample channel region. Note that $\mat{\Sigma}_{L(R)}$ in Eq.~(\ref{eq:Gr2D}) merely represents a matrix representation of $\Sigma_{L(R)}$ in real-space with an orbital and spin basis. The transmission from the left to right contact is obtained from\cite{Datta2005}
\begin{equation} \label{eq:T}
T(E)=\text{Tr}\left[ \mat{G}^r \mat{\Gamma}_L \mat{G}^a \mat{\Gamma}_R \right],
\end{equation}
where $\mat{G}^a=(\mat{G}^r)^\dagger$, and $\mat{\Gamma}_{L(R)}=-2\text{Im}\{ \mat{\Sigma}_{L(R)} \}$.

Fig.~\ref{fig:transport} shows the transmission spectra of the gapless and gapped phase. The dotted line in Fig.~\ref{fig:transport}(a) and (b) shows the transmission spectra as a function of energy in an absence of the spin fluctuation ($\theta_{max}=0$) and serves as a reference. When the AFS is in gapless phase, Fig.~\ref{fig:transport}(a) clearly shows non-zero transmission values near $E=0$ eV.
In contrast, the transmission of the gapped phase is negligible ($<10^{-20}$) within the gap of the AFS.
We then examine 20 different random spin fluctuation configurations to obtain the perturbed N\'{e}el vector configuration, whose averaged N\'{e}el vector orientation is remained as $\hat{\mathbf{n}}||[100]$ for the gapless and $\hat{\mathbf{n}}||[001]$ for the gapped phase. We choose two different $\theta_{max}$ to represent the system with a small fluctuation ($\theta_{max}=\pi/8$) and a large fluctuation ($\theta_{max}=5\pi/8$) in Figs.~\ref{fig:transport}(a) and (b), respectively. In both cases, the transmission spectrum of the gapless phase is in a blue solid line, whereas a red solid line depicts the transmission spectrum of the gapped phase. We first focus on Fig.~\ref{fig:transport}(a) to examine the system response when the fluctuation is small ($\theta_{max}=\pi/8$). Although we observe minute changes in transmission signals for the gapless phase, we observe a small but noticeable overall increase of the transmission spectrum in the gapped phase. More obvious changes in the transmission spectrum of the gapped phase has been observed in Fig.~\ref{fig:transport}(b) with $\theta_{max}=5\pi/8$. We clearly observe that the self-energy of the corresponding fluctuation renormalizes the gap size of the gapped phase, thus observe a subsequent overall increase in transmission values within the gap as well as a decrease in the width of the transmission spectrum dip that corresponds to the gap size. In contrast, the gapless phase still exhibits minimal changes and remains gapless which agrees with our analysis in Section~\ref{sec:renormalize_gapless}.

We may illustrate the gap narrowing effect for the gapped phase by tracking the changes of the minimum transmission values as a function of $\theta_{max}$. Figure~\ref{fig:transport}(c) shows the transmission at a mid-gap energy, $E=E_0=-40$ meV, where we obtain the minimum transmission value for the gapped phase. The spectrum shows a minimum transmission near $E=-40$ meV due to the non-zero $t'_{xy}$ which shifts the location of the Dirac cones in energy. The gray line shows transmission values for 20 different random configurations, and red rectangular and blue circular symbols represent the averaged transmission at $E=E_0$ for the gapped and gapless phase, respectively. Figure~\ref{fig:transport}(c) clearly illustrates an increasing transmission of the gapped phase as a function of $\theta_{max}$ whereas the gapless phase shows no significant changes over entire range of $\theta_{max}$. For the massive Dirac fermion system, the transmission of an electron injected within the mass gap follows\cite{Setare2010, Cao2011} $T\propto e^{-2\kappa d}$, where $d$ is the channel length, $\kappa$ is an evanescent wavevector which satisfies $(\hbar v_F \kappa)^2=E_g^2-E^2$, $E_g$ is a mass gap of the Dirac fermion, and $v_F$ is Fermi velocity. When an electron is injected at the Dirac point ($E=0$ eV), the transmission follows $T\propto e^{-2E_g d/\hbar v_F}$. Equation~(\ref{eq:a5}) shows that the renormalized gap size follows $E_g\propto \cos^2(\theta_{max}/2)$ up to the first order correction, thus we establish a relationship that satisfies $T=f_Be^{-f_A\; \cos^2(\theta_{max}/2)}$, where $f_A$ and $f_B$ are fitting parameters. The dashed line in Fig.~\ref{fig:transport}(c) shows the fitting results which shows a good agreement with the numerically calculated transmission values at $E=E_0$, especially for small $\theta_{max}$. The result confirms that the first order correction in the impurity potential $\mat{V}$ captures the band gap narrowing effect induced by the thermal fluctuation.

\section{Micromagnetics analysis on spatially correlated fluctuation} \label{sec:LLG}

Our previous discussion in Section~\ref{sec:configave} is based on the assumption that spin fluctuations are spatially uncorrelated, thus the impact of the N\'{e}el vector fluctuations are evaluated simply by self-averaging over uncorrelated random N\'eel vector fluctuations. Such approach may be valid for a macroscopic sample whose size is much larger than the phase coherent length of the electron.
However, nanometer size samples may exhibit quantitative difference in their observables as the sample may not contain enough sub-systems to average out the microscopic details such as microscopic spatial correlations in the N\'eel vector fluctuation. 
To properly consider microscopic details, another route to understand the N\'eel vector fluctuation is to treat the microscopic local magnetic moment $\bm{m}_i$ in Eq.~(\ref{eq:int}) as a classical magnetic moment at an atomic site $i$. Assuming that the magnitude of the magnetic moment and the interaction constant are consistent for all the unit cells, the atomistic spin dynamics may be described by the stochastic Landau-Lifshitz-Gilbert (LLG) equation\cite{Garanin1997,Garanin2006, Miyashita2015}. In this micro-spin LLG framework, the coupling of the classical spins with its environment is considered via a phenomenological damping constant $\alpha$. In addition, the white noise type random forces are introduced in LLG equation to describe the thermal fluctuations and the resultant Brownian motion of the local spins at each unit cell\cite{Brown1963, Kubo1970}. An obvious advantage of this approach is that we may properly take account for the microscopic spatial correlations of the classical spins in the presence of thermal fluctuation. In this section, we solve for the spin dynamics and compare the results with that of the self-averaging method in Section~\ref{sec:configave}, and examine the impact of the spatial correlation in spin fluctuations on observables.

\subsection{Micromagnetics model}
To capture the spatial correlation of the spin fluctuation, we solve the micro-spin dynamics in the presence of random forces exerting on the spin at each site. Although each individual spins are perturbed by uncorrelated random white noise, the resultant motion of spin may have a spatial correlation as the spin at each lattice site is affected by the adjacent spins by the exchange coupling.
In this regard, more realistic treatment on the random fluctuation of the N\'{e}el vector is to solve the stochastic Landau-Lifshitz-Gilbert (LLG) equation\cite{Francisco1998,Brown1963,Kubo1970} that captures spin dynamics under the influence of random fluctuations. The equation of motion is\cite{Landau1935,Gilbert1955}
\begin{equation} \label{eq:LLG}
\frac{\partial \bm{m}_i}{\partial t}=
-\gamma \bm{m}_i\times [\mathbf{h}^{eff}_i+\mathbf{h}^{th}_i]
-\gamma\frac{\alpha}{m_s}\bm{m}_i\times (\bm{m}_i\times [\mathbf{h}_{eff}+\mathbf{h}_{th}]),
\end{equation}
where $\bm{m}_i$ is the local magnetization in Eq.~(\ref{eq:int}) at site $i$, $m_s=|\bm{m}_i|$ is the saturation magnetization, $\gamma$ is gyromagnetic ratio, and $\alpha$ is Gilbert damping coefficient. In Eq.~(\ref{eq:LLG}), $\mathbf{h}^{eff}_i=-\delta E_{tot}/\delta \bm{m}_i$ is the effective field where $E_{tot}$ is the total energy of the sample and $\mathbf{h}^{th}_i$ is the random forces describing the thermal fluctuation. Here, the random force is assumed to follow the Gaussian stochastic process which satisfies
\begin{equation}
\langle \mathbf{h}^{th}_i(t)\rangle=0,\quad
\langle \mathbf{h}^{th}_i(t) \mathbf{h}^{th}_j(t') \rangle
=D\delta_{ij}\delta(t-t'),
\end{equation}
where $i,j$ are the lattice indices, $t$ is the time index, and $D$ measures the strength of the thermal fluctuation. According to the fluctuation-dissipation theorem\cite{Landau1980}, the thermal fluctuation field is related to the dissipation of energy, or the damping dynamics of the magnet. Then, the strength of the thermal fluctuation is\cite{Brown1963, Kubo1970, Berkov2007, Bauer2008}
\begin{equation}
D=\frac{\alpha}{1+\alpha^2}\frac{2k_B T}{\gamma \mu_0 m_s V_u},
\end{equation}
where $\mu_0$ is permeability, $V_u$ is a unit cell volume, and $T$ is temperature.

 \begin{figure}[t!]
  \centering
   \includegraphics[width=0.5\textwidth]{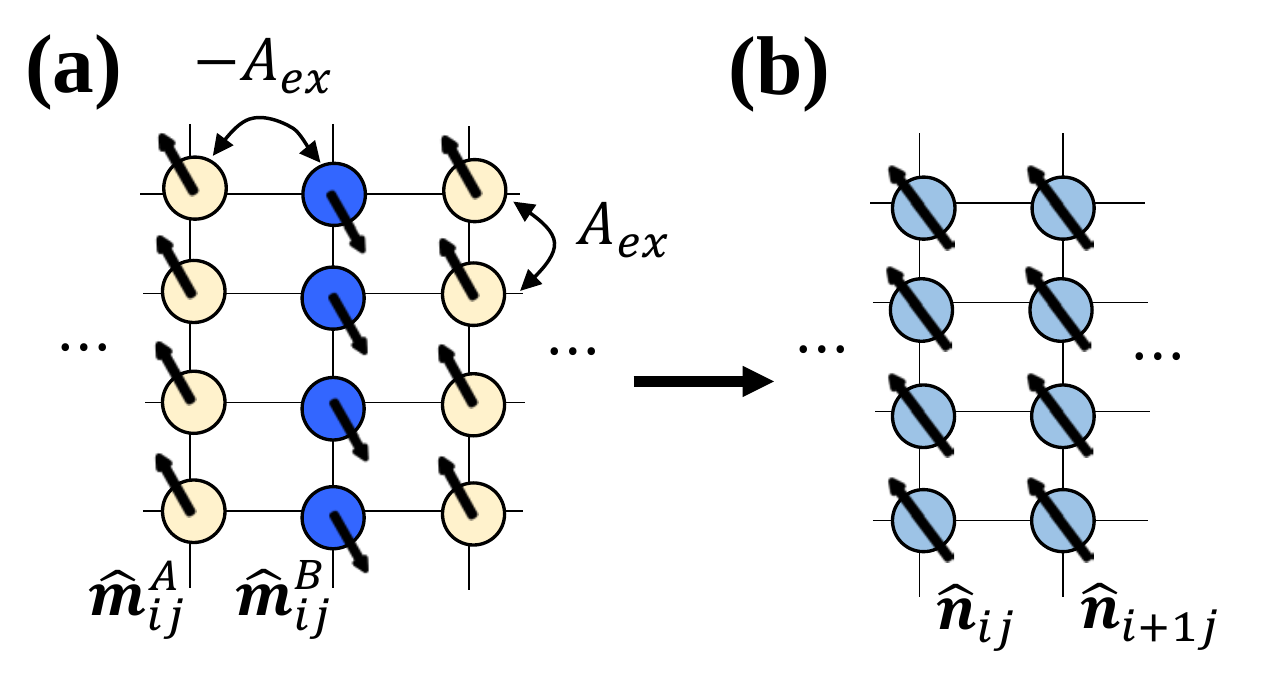}
  \caption{(a) 2D lattice structure under consideration. A and B sublattice structure exhibit staggered magnetic order, together forming an antiferromagnetic order. To maintain an antiferromagnetic order, we introduce exchange coefficient $-A_{ex}$ between A and B sublattices. Meanwhile, we use $A_{ex}$ between the same sublattices in order to maintain ferromagnetic order. (b) We form a N\'{e}el vector $\dvec{n}$ by post-processing micromagnetics simulation results, and the resultant unit cell structure is shown in green circle. The configuration is imported as an imput for NEGF calculation.
  }\label{fig:schematics2}
\end{figure}   

Fig.~\ref{fig:schematics2}(a) shows the lattice structure used for the micromagnetic simulations. The system consists of $A$ and $B$ sublattice degree of freedom each of which possesses a staggered magnetization orientation, together forming the N\'{e}el order. In a given rectangular unit cell structure, the exchange energy is calculated as\cite{Donahue1999}
\begin{equation} \label{eq:Eex}
\begin{split}
E^A_r=&
  \sum_{k\in \{i-1,i\}}(-A_{ex})\frac{\bm{m}^A_{ij}\cdot (\bm{m}^A_{ij}-\bm{m}^B_{kj})}{\Delta x^2} \\
+&\sum_{l\in \{j-1,j+1\}}A_{ex}\frac{\bm{m}^A_{ij}\cdot (\bm{m}^A_{ij}-\bm{m}^A_{il})}{\Delta x^2}, \\
E^B_r=&
  \sum_{k\in \{i,i+1\}}(-A_{ex})\frac{\bm{m}^B_{ij}\cdot (\bm{m}^B_{ij}-\bm{m}^A_{kj})}{\Delta x^2} \\
+&\sum_{l\in \{j-1,j+1\}}A_{ex}\frac{\bm{m}^B_{ij}\cdot (\bm{m}^B_{ij}-\bm{m}^B_{il})}{\Delta x^2}, \\
\end{split}
\end{equation}
where $\bm{m}_{r}^{A/B}=\bm{m}_{ij}^{A/B}$ is the local magnetization with the unit cell index $r=(i,j)$ that describes the $x$ and $y$ directional lattice site indices, $A_{ex}$ is the exchange coefficient between the nearest neighbor sites, and $\Delta x$ is the discretized step size. The exchange energy in Eq.~(\ref{eq:Eex}) is then summed up to constitute the total energy, $E_{tot}=\sum_r (E_{r}^A+E_{r}^B)+\cdots$. To maintain an antiferromagnetic order, we use an exchange coefficient $-A_{ex}$ between $A$ and $B$ sublattices and $A_{ex}$ within the same sublattices. Once the spin dynamics are calculated, the N\'{e}el vector configuration at each site is computed from
\begin{equation} \label{eq:nij}
\dvec{n}_{r}=\frac{1}{2}(\dvec{m}_{r}^A-\dvec{m}_{r}^B),
\end{equation}
where $\dvec{m}_{r}^{A/B}=\bm{m}_{r}^{A/B}/|\bm{m}_{r}^{A/B}|$ is the orientation of local magnetization at site $\mathbf{r}=(i,j)$.
Fig.~\ref{fig:schematics2}(b) shows the post processed N\'eel vector orientation $\dvec{n}_{ij}$. Then the N\'{e}el vector orientation at each site is fed to the antiferromagnetic interaction Hamiltonian in Eq.~(\ref{eq:intAF}).
In addition, a deviation from staggered order of $A$ and $B$ sublattices may induce a ferromagnetic order $\dvec{l}_{\mathbf{r}}=\frac{1}{2}(\dvec{m}_{i}^A+\dvec{m}_{i}^B)$. Within the mean-field approximation, the ferromagnetic order may be included in the Hamiltonian as $\dvec{l}\cdot\bm{\sigma}$. Then such term breaks underlying symmetries and may induce a gap proportional to the magnitude of the ferromagnetic order. However, we assume that the antiferromagnetic exchange energy is sufficiently strong to maintain the antiferromagnetic order and ignore $\dvec{l}$ for our remaining discussions.

We now aim to evaluate the impact of thermal fluctuation on device transport characteristics. Specifically, our main goal is to evaluate how the thermal fluctuation obtained from micromagnetics simulation manifests as a deviation from transport characteristics of the ideal system. To this end, we first solve the stochastic LLG equation\footnote{Numerical calculation has been performed by using open source software OOMMF\cite{Donahue1999}.} to obtain the N\'{e}el vector evolution up to the total simulation time of $6$ ns on $N_x\times N_y$ unit cells where $N_x$ and $N_y$ are the number of unit cells in width and length direction, respectively.
We define an uniaxial anisotropy energy $K_u$, whose direction is assumed to be in $[100]$ for the gapless and $[001]$ for the gapped phase.
We sample 20 points out of an interval from $2$ ns to $6$ ns simulation time to avoid any bias in the N\'{e}el vector configuration by initialization of simulations. The sampled spin configurations are assumed to represent a possible N\'{e}el vector configuration and we calculate transport characteristics using NEGF formalism for each spin configuration to obtain an ensemble average of the current signal. Due to the stochastic nature of the thermal fluctuation, we repeat the same procedures for 10 different random seeds and obtain the final current value by averaging the results.

For each N\'{e}el vector configuration, we construct the retarded Green function in Eq.~(\ref{eq:Gr2D}) and compute the transmission, $T(E)$, by using Eq.~(\ref{eq:T}). Finally, the resultant current is calculated as\cite{Datta2005}
\begin{equation} \label{eq:current}
I=\frac{2e}{h}\int_{-E_{cut}}^{E_{cut}}\;\frac{dE}{2\pi}\;T(E)[f_L(E)-f_R(E)],
\end{equation}
where $f_{L(R)}(E)=1/(1+e^{(E-\mu_{L(R)})/k_B T})$ is Fermi-Dirac distribution function of the left (right) contact at temperature $T=300$ K with the contact chemical potential $\mu_{L(R)}$, $k_B$ is Boltzmann constant, $h$ is Plank's constant, and $E_{cut}$ is a numerical energy integration cutoff. In this calculation, we set $E_{cut}=0.1\text{ eV}\simeq4k_BT$ to sufficiently capture the thermal electron contributions, and we apply a small bias $\mu_L-\mu_R=10$ meV to induce a net current flow across the sample.

 \begin{figure}[t]
  \centering
   \includegraphics[width=0.4\textwidth]{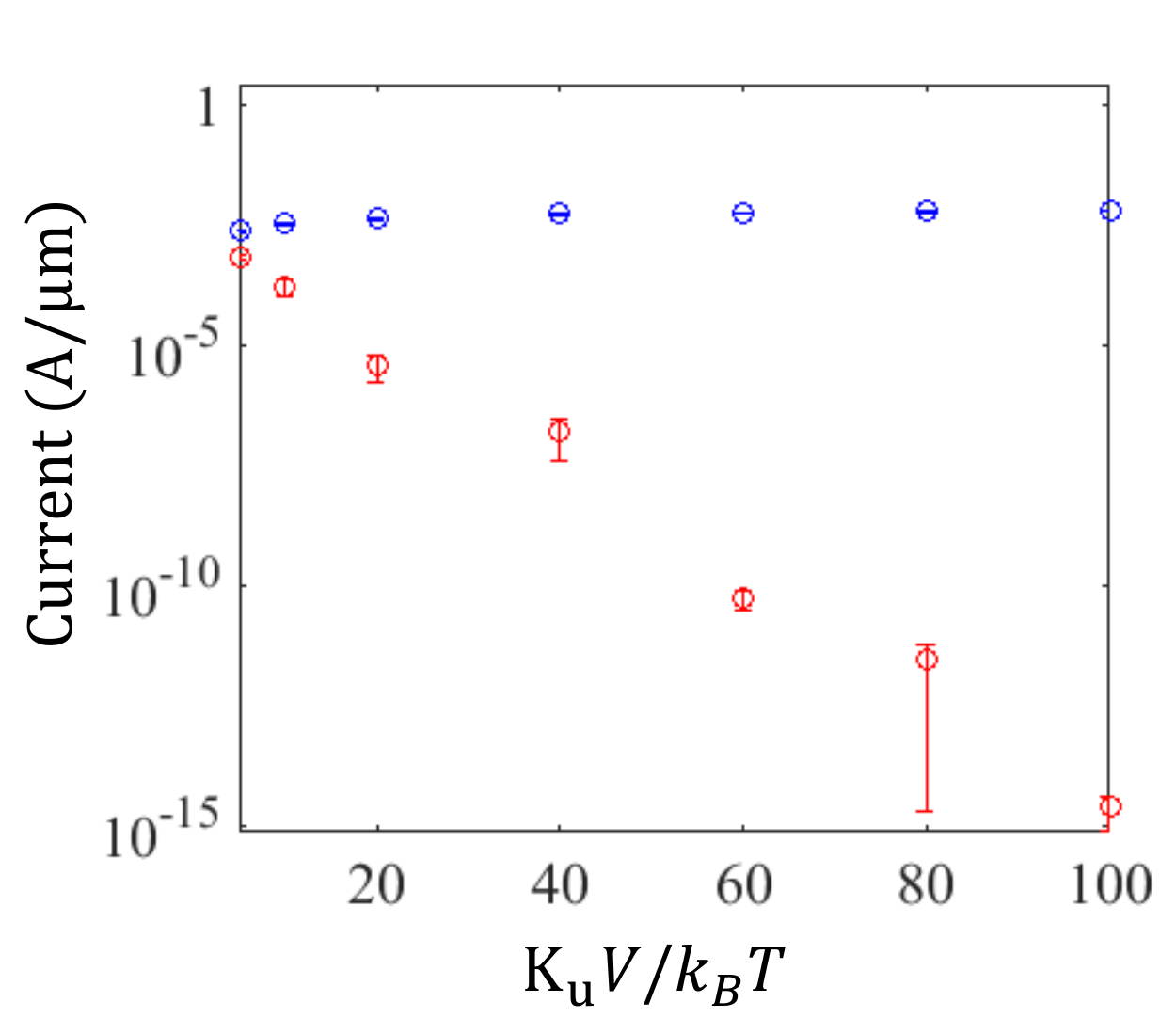}
  \caption{
The two terminal current of AFS device with spatially correlated random thermal fluctuation model. We directly import the N\'{e}el vector configurations from micromagnetics simulations and each ensemble average current is obtained for 20 different snapshots uniformly sampled from $2$ ns to $6$ ns spin evolution time. The same procedures are repeated over 10 different random seeds, and the resultant averaged current is presented with an standard error.
The red circular symbol represents the current from the gapped phase ($\dvec{n}||[001]$), and the blue rectangular symbol shows the current of the gapless phase ($\dvec{n}|[100]|$). The thermal barrier $U_{th}=K_uV/k_B T$ is adjusted in order to effectively tune the maximum fluctuation angle in micromagnetics simulation. The red circular symbol represents the current from the gapped phase ($\dvec{n}||[001]$), and the blue rectangular symbol shows the current of the gapless phase ($\dvec{n}|[100]|$).
We use following micromagetics simulation parameters: $\Delta x=1$ nm, $m_s=800$ kA/m, $A_{ex}=10$ pJ/m, $\alpha=0.01$, and $T=300$ K. We use the same NEGF parameters used in Fig.~\ref{fig:transport}, and the device for both (a) and (b) is $N_x\times N_y=50\times 100$.
  }\label{fig:fluct_KuV}
\end{figure}   
\subsection{Thermal barrier and transport gap narrowing}
Once we perform the micromagnetics simulation and allow spin dynamics to determine the thermal fluctuations, it is not possible to manually adjust any phenomenological parameters such as $\theta_{max}$ to tune the magnitude of the fluctuation. Instead, we adjust the thermal barrier\cite{Chshiev2017}, $U_{th}=K_u V/k_B T$, which is defined in Eq.~(\ref{eq:Uth}). We allow large averaged thermal fluctuation angle for a given temperature by lowering the anisotropy energy $K_u$ to reduce the thermal barrier of the system. Fig.~\ref{fig:fluct_KuV} shows the averaged current of $N_x\times N_y=50\times100$ lattice system for $200$ different random configurations (10 different random seeds for thermal fluctuation, 20 sampled configurations from each random seed), whose N\'{e}el vector configurations are obtained from the micromagnetics simulation. For decreasing $U_{th}$, we observe that the current of the gapped phase substantially increases. The observed qualitative behavior agrees with that of the Fig.~\ref{fig:transport}(c) where the larger $\theta_{max}$ results in a smaller gap size in the gapped phase, thereby increases the tunneling current between two contacts. Moreover, the current of the gapless phase is relatively unaffected, which also agrees with the qualitative behavior found in Fig.~\ref{fig:transport}(c). Therefore, we numerically confirm that the qualitative behavior of the current in the presence of microscopic N\'{e}el vector fluctuation may still be captured and explained by the self-averaging method presented in Section~\ref{sec:configave}.

\subsection{The spatially correlated fluctuation and its impact on two-terminal current}
 \begin{figure*}
  \centering
   \includegraphics[width=1.0\textwidth]{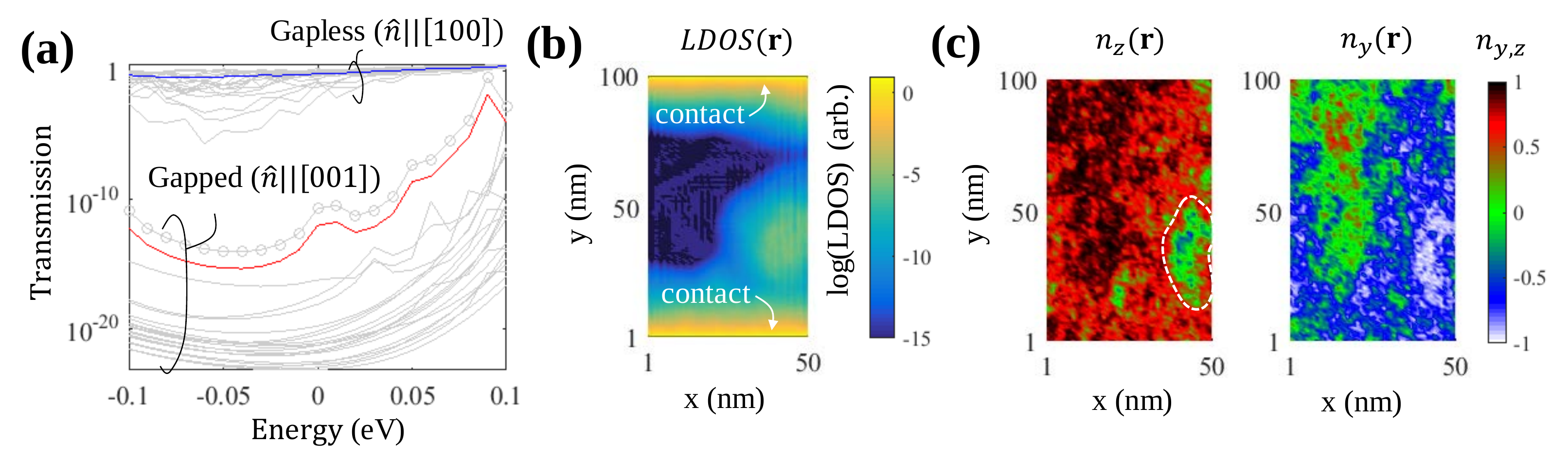}
  \caption{(a) The transmission plot as a function of energy for $L_x\times L_y=50\times 50$ nm$^2$ system at one specific random seed. The gray line shows 20 different transmission plots which corresponds to the different N\'{e}el vector configurations uniformly sampled from $2$ ns to $6$ ns spin evolution. The averaged transmission value for gapped and gapless phase are indicated as red and blue lines, respectively. The plot emphasize that the averaged transmission for gapped phase is dominantly determined by the one random configuration which shows abnormally high transmission plot indicated in circular gray symbol. To examine more details of this particular configuration, we plot spatially resolved local density of states (LDOS) in (b). (b) The plot of LDOS that corresponds to the N\'{e}el vector configuration which produces circular gray symbol in (a). The LDOS is plotted for $E=0$ eV, and its magnitude is plotted in log scale. (c) The corresponding N\'{e}el vector configuration. Here, we only present $\hat{z}$ and $\hat{y}$ directional N\'{e}el vector components in left and right side, respectively. The majority lattice site has N\'{e}el vector aligned with $\hat{z}$ direction, as the system is initially in gapped phase. Meanwhile, we observe puddles in leftmost side of the device, whose net N\'{e}el vector is oriented in negative $\hat{y}$ direction (highlighted in white dashed line). The puddles locally closes the gap and provide available states even at $E=0$ eV as shown in LDOS in (b).
  }\label{fig:fluct_snapshot}
\end{figure*}
However, a close analysis on each point in Fig.~\ref{fig:fluct_KuV} reveals that we still may need to consider microscopic details when it comes to nanometer size samples. To begin with, we take a closer look in one particular fluctuation configuration in Fig.~\ref{fig:fluct_KuV}. Figure~\ref{fig:fluct_snapshot}(a) shows a transmission of the device with $N_x\times N_y=50\times 100$ at $U_{th}=k_uV/k_B T=40$. The result shows 20 different snapshots of the N\'{e}el vector configurations for a given random seed, and the resultant transmission spectra are presented in gray lines in Fig.~\ref{fig:fluct_snapshot}(a) both for the gapped and gapless phase. The averaged value is indicated as a red (blue) solid line for the gapped (gapless) phase. Two out of 20 different configurations show noticeably high transmission values for the gapped phase, which dominantly determine the resultant tunneling current of the sample. We select a particular N\'{e}el vector configuration that produces the highest transmission in the gapped phase (gray circular symbol) in Fig.~\ref{fig:fluct_snapshot}(a) and plot the spatially resolved local density of states (LDOS) in Fig.~\ref{fig:fluct_snapshot}(b) at $E=0$ eV. The source and drain contacts are connected to the top and bottom side of the sample. Here, we observe relatively high DOS near the contacts due to the wavefunction penetration into the sample channel region. Due to the fact that the energy of interest lies inside of the gap, the channel region has no available states showing an exponential decay of LDOS from contact to center. However, Fig.~\ref{fig:fluct_snapshot}(b) shows finite LDOS on the right-hand side of the device which provides a conducting channel and serves as a source of the enhanced transmission indicated as circular symbols in Fig.~\ref{fig:fluct_snapshot}(a).
To further clarify the origin of the enchanced tunneling current source, Fig.~\ref{fig:fluct_snapshot}(c) shows the spatially resolved local N\'{e}el vector configurations in $\hat{z}$ and $\hat{y}$ direction when the sample is in gapped phase. Due to the uniaxial anisotropy in $[001]$ direction, most of the N\'{e}el vectors are aligned in $[001]$ direction and, thus the channel is insulating. However, we observe a spatially correlated local fluctuations that forms a puddle of local N\'{e}el vectors aligned in $\hat{y}$ direction (white dashed circle). The puddle that is directed toward $\hat{y}$ significantly reduces a local gap size and may provide available states. When such event happens near the contact region, the puddle effectively shortens the channel length. The tunneling current has exponential dependency on the length of the barrier\cite{Griffiths2016}, therefore, shortening the length of the insulating channel leads to an exponential increase in the tunneling current. As a result, the current in the gapped phase of the microscopic sample may be dominantly determined by the rare event where the puddles are formed in a way to effectively shorten the insulating barrier length of the channel.

However, this is not the case for the macroscopic sample. Here, the size of the puddle spans few nm, as the exchange length\cite{Choi2013} of our system is $l_{ex}=\sqrt{2A_{ex}/\mu_0 m_s^2}\simeq5$ nm. The system whose size is much larger than $l_{ex}$ is less likely to be affected by puddles, as maximum modification of $\sim l_{ex}$ in channel length is negligible for the macroscopic sample. In addition, it is unlikely to have series of connected isolated puddles that induces local conducting paths in a specific direction. In this regard, our analysis in this section may be valid to samples whose channel length is comparable to the correlation length of the N\'eel vector fluctuations (e.g. $l_{ex}$).

The same scenario may occur to the gapless phase of the AFS. For example, the spatially correlated N\'eel vector fluctuation puddle may divert the local N\'eel vector from $\hat{x}$ to $\hat{z}$ direction. Then the corresponding sites may have a local gap as the underlying symmetry is locally broken. Although such insulating puddle may decrease the overall conductivity of the sample, such gap is still local and remaining channel regions provide conducting paths maintaining its semimetallic phase. Therefore, the impact of such puddle is less significant compared with that of the gapped phase.

\section{Summary and conclusion} \label{sec:conclusion}
As the thermal fluctuation always exists in antiferromagnetic semimetals (AFS), it may be viewed as an \emph{intrinsic} disorder. In this regard, it is essential to understand the role of the thermal fluctuation in AFS and its impact on observables.

We first study the macroscopic impact of the thermal fluctuation in the N\'{e}el vector orientation on the quasi-particle spectrum of AFS. By treating the thermal fluctuation as a spatially uncorrelated impurity potential, we examine the renormalized Hamiltonian by self-averaging over all the possible fluctuations. As a result, we find that the symmetry protected Dirac semimetal phase is relatively robust to the N\'eel vector fluctuation.
When the underlying symmetry is broken and the spectrum is gapped, however, the size of the gap is reduced for the increasing thermal fluctuation. We numerically confirm the results by using the real-space Hamiltonian and the non-equilibrium Green function formalism.

In microscopic samples, however, spin dynamics are spatially correlated and spin fluctuations form puddles in real-space rather than manifest itself as an uncorrelated white noise. Assuming classical magnetic moments at each atomic sites, we solve the stochastic Landau-Lifshitz-Gilbert (LLG) equation and examine the effect of such fluctuation puddles for microscopic samples. We find that the fluctuation puddles induce a local phase transition and may provide available states in the channel even in the case when the sample is initially in the gapped phase. Especially, when such available states are formed near the contact region, the effective channel length becomes shorter providing an exponentially enhanced tunneling current from one contact to the other. Consequently, we no longer have a perfect insulating phase and may obtain significantly degraded anisotropy magnetoresistance ratio (AMR) in nanometer size AFS samples.

The relationship of the thermal fluctuation and the gap size of the AFS may be identified via the optical conductivity measurement. The linear relationship between the optical conductivity and the frequency has been utilized to identify the Dirac semimetal phase\cite{Attila2013,Pavan2012,Wang2015,Chen2017,Conte2017}. In addition, a sudden deviation from the linear relationship near the Dirac point serves as a signature of a gap in Dirac semimetals\cite{Chen2017}. When AFS is in the insulating phase, one may utilize the measured optical conductivity to quantify the gap size for an increasing temperature, thereby identify the reduction of the gap at elevated temperatures.

Our results provide a comprehensive understanding on the possible impact of thermal fluctuations in AFS. We hope that our results serve as a useful guideline in understanding the semimetal-insulator transition in AFS at elevated temperatures.

\begin{acknowledgements}
This work is supported by the National Science Foundation (NSF) under Grant No. DMR-1720633. Y. Kim acknowledges useful discussions from T. M. Philip. Y. Kim also would like to show a gratitude to A. Hoffmann for sharing his initial concerns and interests on the thermal fluctuation in antiferromagnetic semimetals. 
\end{acknowledgements}

\appendix

\begin{widetext}

\section{Spin-spin interaction Hamiltonian} \label{app:Hint}
The interaction Hamiltonian in Eq.~(\ref{eq:int}) is obtained by considering the tight-binding model by considering the itinerant electrons and local magnetic moment from localized orbitals\cite{Sinova2016}. However, similar equaiton may be deduced by taking account for the local exchange interaction between itinerant electrons in $A$ and $B$ sublattices\cite{Vleck1941, Tserkovnyak2018}. The spin-spin interaction Hamiltonian is given as
\begin{equation} \label{eq:Hint}
H_{int}=J_1\sum_{\langle i j\rangle} \mathbf{s}_i\cdot \mathbf{s}_j
-J_2\sum_{\langle\langle i j\rangle\rangle} \mathbf{s}_i\cdot \mathbf{s}_j,
\end{equation}
where $\mathbf{s}_i=c_i^\dagger \bm{\sigma} c_i$ is the spin density operator, $J_1>0$ is the interaction constant between $A$ and $B$ sublattice, and $J_2>0$ is the interaction constant within the same sublattice. In Eq.~(\ref{eq:Hint}), the sign in front of the interaction constant is chosen to reflect the specific magnetic orders illustrated in Fig.~\ref{fig:schematics2}(a) where an antiferromagnetic order is established between $A$ and $B$ sublattice sites whereas a ferromagnetic order is formed within the same sublattices. To obtain a single-particle Hamiltonian, we utilize an arbitrary localized Wannier function (WF) at site $i$, or $\ket{i}$, to project the spin densities on the local sites at $\ket{i}$, or $\mathbf{s}_i=\bra{i}\mathbf{s}_i\ket{i}+\delta\mathbf{s}$, where $\delta\bm{s}$ is a residual spin density reflecting the difference between the true spin density with the projected spin density. Assuming a small residual spin density, or $\delta\bm{s}\ll 1$, we follow the standard procedures of the mean-field decomposition and drop the terms containing a second order in $\delta\bm{s}$. As a result, we obtain the single-particle interaction Hamiltonian\cite{Tserkovnyak2018}
\begin{equation} \label{eq:HintWF}
\begin{split}
H_{int}\simeq H_{int}^{WF}=&
\sum_{i} (J_1\bm{\bar{S}}_{1,i}-J_2 \bm{\bar{S}}_{2,i})\cdot \bm{s}_i
\end{split}
\end{equation}
where $\bm{\bar{S}}_{1,i}=\sum_{\langle ij\rangle} \bra{j}\bm{s}_j\ket{j}$ is the total projected local spin density of the nearest-neighbor sites ($A$-$B$), and $\bm{\bar{S}}_{2,i}=\sum_{\langle\langle ij\rangle\rangle} \bra{j}\mathbf{s}_j\ket{j}$ is the total projected local spin density of the next-nearest-neighbor sites ($A$-$A$, or $B$-$B$). 
To further simplify Eq.~(\ref{eq:HintWF}), we define a unit cell consists of adjacent $A$ and $B$ sublattices at site $i$ and $i+1$, respectively, for $i\in\{1,2,\cdots,2N\}$, where $2N$ is the total number of atomic sites. Then, we index each unit cell location as $r\in\{1,2,\cdots,N\}$. Within each unit cell, the spin density operator at site $A$ and $B$ are labeled as $\bm{s}_{r}^A$ and $\bm{s}_{r}^B$, respectively, and the corresponding projected local spin density is symbolized as $\bm{\bar{S}}_{1(2),r}^{A/B}$. Therefore, we rewrite Eq.~(\ref{eq:HintWF}) as
\begin{equation} \label{eq:HintWF2}
\begin{split}
H_{int}^{WF}=&
\sum_{r} (J_1\bm{\bar{S}}_{1,r}^A-J_2 \bm{\bar{S}}_{2,r}^A)\cdot \bm{s}_i^A
+(J_1\bm{\bar{S}}_{1,r}^B-J_2 \bm{\bar{S}}_{2,r}^B)\cdot \bm{s}_i^B.
\end{split}
\end{equation}
Further simplification can be made by assuming that $\bm{\bar{S}}_{1(2),r}^A$ and $\bm{\bar{S}}_{1(2),r}^B$ have a perfect antiferromagnetic order and satisfy
\begin{equation} \label{eq:AF}
\begin{split}
\bm{\bar{S}}_{1,r}^A=&\bm{\bar{N}}_{1,r},\;\bm{\bar{S}}_{1,r}^B=-\bm{\bar{N}}_{1,r},\\
\bm{\bar{S}}_{2,r}^A=&-\bm{\bar{N}}_{2,r},\;\bm{\bar{S}}_{2,r}^B=\bm{\bar{N}}_{2,r},
\end{split}
\end{equation}
where $\bm{\bar{N}}_{1,r}$ and $\bm{\bar{N}}_{2,r}$ are locally projected nearest-neighbor and next-nearest-neighbor spin density at $r$th unit cell with an assumption that the magnetic moment at A and B sublattices are anti-aligned. Note that $\bm{\bar{N}}_{1,r}$ and $\bm{\bar{N}}_{2,r}$ are aligned each other, and the specific sign choice in Eq.~(\ref{eq:AF}) is due to the fact that $\bm{\bar{S}}_{1,r}^A$ is the local spin density projected to the adjacent $B$ sublattice sites whereas $\bm{\bar{S}}_{2,r}^A$ is the spin density projected to the adjacent $A$ sublattice sites, thereby having an opposite sign. Then, Eq.~(\ref{eq:HintWF2}) becomes
\begin{equation} \label{eq:HintWF3}
H_{int}\simeq H_{int}^{AF}=\sum_{r} \mathbf{\Delta}_r \cdot [\mathbf{s}_{r}^A - \mathbf{s}_{r}^B]
=\sum_{r} \Delta_r \bm{\hat{n}}_r\cdot [\mathbf{s}_{r}^A - \mathbf{s}_{r}^B],
\end{equation}
where we define $\mathbf{\Delta}_r=J_1\bm{\bar{N}}_{1,r}+J_2\bm{\bar{N}}_{2,r}=\Delta_r\bm{\hat{n}}_r$ and $\vert \mathbf{\Delta}_r\vert=\Delta_r$. In Eq.~(\ref{eq:HintWF3}), $\bm{\hat{n}}_r$ indicates the orientation of the projected spin angular momentum. As a result, we reproduce the approximated single-particle antiferromagnetic order interaction Hamiltonian in Eq.~(\ref{eq:intAF}) in the main text.

\section{Hamiltonian and eigenstates} \label{sec:Ham_and_eig}
The tight-binding model Hamiltonian for antiferromagnetic semimetal in Eq.~(\ref{eq:H3D}) is simplified by using parameters $a_{0,1,2,3,4,5}$ in Eqs.~(\ref{eq:Hsimple}), (\ref{eq:a012345}).
The eigenvalues and eigenstates for Eq.~(\ref{eq:Hsimple}) are given as

\begin{equation} \label{eq:UD}
\begin{split}
U=&\frac{1}{\sqrt{2a}}
\begin{pmatrix}
\frac{1}{\sqrt{a+a_5}}
\begin{pmatrix}
a_3-ia_4 & -a_1+ia_2 \\
-a_5-a & 0 \\
0 & a_5+a \\
a_1+ia_2 & a_3+ia_4 \\
\end{pmatrix} &
\frac{1}{\sqrt{a-a_5}}
\begin{pmatrix}
a_3-ia_4 & -a_1+ia_2 \\
-a_5+a & 0 \\
0 & a_5-a \\
a_1+ia_2 & a_3+ia_4 \\
\end{pmatrix} \\
\end{pmatrix} \\
D=&a_0 +
\begin{pmatrix}
-a &  &  & \\
 & -a &  & \\
 &  & a & \\
 &  &  & a \\
\end{pmatrix},
\end{split}
\end{equation}
respectively, where $a=\sqrt{a_1^2+a_2^2 + a_3^2+a_4^2+a_5^2}$.
Knowing the eigenstates, we obtain Green function from following procedure:
 \begin{equation} \label{eq:Gappendix}
 \begin{split}
 G^{-1}=&(\epsilon-H+i0^+)
 =UU^\dagger (\epsilon-H+i0^+) UU^\dagger \\
 =&U (\epsilon-U^\dagger HU+i0^+) U^\dagger
 =U (\epsilon-D+i0^+) U^\dagger, \\
 G=& U (\epsilon-D+i0^+)^{-1} U^\dagger = G_- + G_+ \\
 =&\frac{1}{\epsilon-(a_0-a)+i0^+}
 U\left(\frac{I+\tau_3}{2}\right)U^\dagger
 +\frac{1}{\epsilon-(a_0+a)+i0^+}
 U\left(\frac{I-\tau_3}{2}\right)U^\dagger.
 \end{split}
 \end{equation}
 After some algebra, we notice that
 \begin{equation} \label{eq:U}
 U\left(\frac{I\pm\tau_3}{2}\right)U^\dagger
 =
 \frac{1}{2}\left(
 1\mp \frac{1}{a}
 (a_1\tau_1
+a_2\tau_2
+ a_3\tau_3\sigma_1
+a_4\tau_3\sigma_2
+a_5\tau_3\sigma_3)
 \right).
 \end{equation}
 As a result, we have Green function for two doubly degenerate states as
 \begin{equation}  \label{eq:Gpm}
 \begin{split}
 G_-=&\frac{1}{\epsilon-(a_0-a)+i0^+}
 \frac{1}{2}\left(
 1- \frac{1}{a}
 (a_1\tau_1
+a_2\tau_2
+ a_3\tau_3\sigma_1
+a_4\tau_3\sigma_2
+a_5\tau_3\sigma_3)
 \right),
   \\
G_+=&\frac{1}{\epsilon-(a_0+a)+i0^+}
 \frac{1}{2}\left(
 1+ \frac{1}{a}
 (a_1\tau_1
+a_2\tau_2
+ a_3\tau_3\sigma_1
+a_4\tau_3\sigma_2
+a_5\tau_3\sigma_3)
 \right). \\
 \end{split}
 \end{equation}

\section{Impurity potential and its self-energy}
\subsection{Transfer matrix}
In the presence of the impurity potential operator, $\mat{V}$, the Green function can be written by using Dyson equation\cite{Sheng1995}
\begin{equation} \label{eq:G1}
\mat{G}=\mat{G}_0 + \mat{G}_0\mat{V}\mat{G},
\end{equation}
where $\mat{G}_0$ is bare Green function. Equation~(\ref{eq:G1}) is obtained by iterating on $\mat{G}$ as
\begin{equation}
\begin{split}
\mat{G}=&\mat{G}_0+\mat{G}_0\mat{V}[\mat{G}_0+\mat{G}_0\mat{V}(\mat{G}_0+\mat{G}_0\mat{V} \dots )]  \\
=& \mat{G}_0+\mat{G}_0\mat{V}\mat{G}_0
+\mat{G}_0\mat{V}\mat{G}_0\mat{V}\mat{G}_0 + \dots \\
=& \mat{G}_0+\mat{G}_0\mat{T}\mat{G}_0,
\end{split}
\end{equation}
where
\begin{equation} \label{eq:T1}
\begin{split}
\mat{T}=&\mat{V}+\mat{V}\mat{G}_0\mat{V}
+\mat{V}\mat{G}_0\mat{V}\mat{G}_0\mat{V}+\dots \\
=&\mat{V}(\mat{I}-\mat{G}_0\mat{V})^{-1}=(\mat{I}-\mat{G}_0\mat{V})^{-1}\mat{V}.
\end{split}
\end{equation}
Here, the $\mat{T}$ matrix is called the scattering matrix. Also, from Eq~(\ref{eq:G1}),
\begin{equation} \label{eq:Ginv1}
\mat{G}^{-1}=\mat{G}^{-1}_0(\mat{I}-\mat{G}_0\mat{V})=\mat{G}_0^{-1}-\mat{V}.
\end{equation}

Let us assume that the impurity potential occurs in a random manner, and we may take a configuraitonal average, which results
\begin{equation} \label{eq:G2}
\braket{\mat{G}}=\mat{G}_0+\mat{G}_0\braket{\mat{T}}\mat{G}_0,
\end{equation}
and
\begin{equation} \label{eq:T2}
\braket{\mat{T}}=\braket{\mat{V}(\mat{I}-\mat{G}_0\mat{V})^{-1}}=\int dv D(v)\mat{V}(\mat{I}-\mat{G}_0\mat{V})^{-1},
\end{equation}
where $\braket{\cdot}$ represents an average over all available impurity configurations, $v$ is a variable for a specific defect configuration, and $D(v)$ is the corresponding defect configuration probability distribution function which satisfies $\int dv D(v)=1$. Analogous to the Eq.~(\ref{eq:Ginv1}), we introduce the self-energy as
\begin{equation} \label{eq:sigma1}
\braket{\mat{G}}^{-1}=\mat{G}_0^{-1}-\mat{\Sigma}.
\end{equation}
To get a closed form for $\Sigma$, we begin with Eq~(\ref{eq:G2}):
\begin{equation} \label{eq:G3}
\begin{split}
\braket{\mat{G}}=
(\mat{I}+\mat{G}_0\braket{\mat{T}})\mat{G}_0
\rightarrow
\braket{\mat{G}}^{-1}=
(\mat{I}+\mat{G}_0\braket{\mat{T}})^{-1}\mat{G}_0 ^{-1}
=
\mat{G}_0 ^{-1}(\mat{I}+\mat{G}_0\braket{\mat{T}})^{-1}.
\end{split}
\end{equation}
Plugging Eq.~(\ref{eq:G3}) into Eq.~(\ref{eq:sigma1}),
\begin{equation} \label{eq:Sigma}
\begin{split}
\mat{\Sigma}=&\mat{G}_0^{-1}-\braket{\mat{G}}^{-1} \\
=&\mat{G}_0^{-1}-\mat{G}_0 ^{-1}(\mat{I}+\mat{G}_0\braket{\mat{T}})^{-1} \\
=&\mat{G}_0 ^{-1}[\mat{I}-(\mat{I}+\mat{G}_0\braket{\mat{T}})^{-1}] \\
=&\mat{G}_0 ^{-1}[\mat{I}+\mat{G}_0\braket{\mat{T}}-\mat{I}](\mat{I}+\mat{G}_0\braket{\mat{T}})^{-1} \\
=&\mat{G}_0 ^{-1}[\mat{G}_0\braket{\mat{T}}](\mat{I}+\mat{G}_0\braket{\mat{T}})^{-1} \\
=&\braket{\mat{T}}(\mat{I}+\mat{G}_0\braket{\mat{T}})^{-1}. \\
\end{split}
\end{equation}

\subsection{Effective medium theory} \label{sec:effective_medium}
Another way of solving this problem is to introduce a concept of \emph{effective medium}, where the scattering sources are renormalized to a form of $\mat{G}_{eff}$. In this perspective, we introduce the effective self-energy, which represents the effective contribution of the impurity potential. To explicitly show this, following transform occurs:
\begin{equation}
\begin{split}
[\epsilon-H_0-\mat{V}] &\rightarrow [\epsilon-H_0-\mat{\Sigma}_{eff}-(\mat{V}-\mat{\Sigma}_{eff})], \\
\mat{G}_0=[\epsilon-H_0]^{-1} &\rightarrow \mat{G}_{eff}=[\epsilon-H_0-\mat{\Sigma}_{eff}]^{-1}, \\
\mat{V} &\rightarrow  \mat{V}-\mat\Sigma_{eff} \\
\braket{\mat{T}} &\rightarrow \braket{\mat{T}_{eff}}=0. \\
\end{split}
\end{equation}
Consequently, Eq.~(\ref{eq:T2}) becomes
\begin{equation} \label{eq:Tinteff}
\braket{\mat{T}_{eff}}=\int dv D(v)[\mat{V}-\mat{\Sigma}_{eff}](\mat{I}-\mat{G}_{eff}[\mat{V}-\mat{\Sigma}_{eff}])^{-1}=0.
\end{equation}
Manipulating Eq.~(\ref{eq:Tinteff}),
\begin{equation}
\begin{split}
 [\mat{V}-\mat{\Sigma}_{eff}](\mat{I}-\mat{G}_{eff}[\mat{V}-\mat{\Sigma}_{eff}])^{-1}
=&
\mat{G}_{eff}^{-1}(\mat{G}_{eff}[\mat{V}-\mat{\Sigma}_{eff}] +\mat{I}-\mat{I})(\mat{I}-\mat{G}_{eff}[\mat{V}-\mat{\Sigma}_{eff}])^{-1} \\
=&
-\mat{G}_{eff}^{-1}
[\mat{I}-(\mat{I}-\mat{G}_{eff}[\mat{V}-\mat{\Sigma}_{eff}])^{-1}].
\end{split}
\end{equation}
As a result,
\begin{equation} \label{eq:TDs2}
\int dv D(v)(\mat{I}-\mat{G}_{eff}[\mat{V}-\mat{\Sigma}_{eff}])^{-1}=\mathbf{I}.
\end{equation}
By plugging in Eq~(\ref{eq:TDs2}) into Eq.~(\ref{eq:Tinteff}),
\begin{equation} \label{eq:Sigeff}
\mat{\Sigma}_{eff}=\int dv D(v)\mat{V}(\mat{I}-\mat{G}_{eff}[\mat{V}-\mat{\Sigma}_{eff}])^{-1}.
\end{equation}
By solving Eq.~(\ref{eq:Sigeff}) self-consistently, we obtain the self-energy of the effective medium.

\section{Evaluating operators} \label{sec:evaluation}
Having Green function in Eq.~(\ref{eq:Gappendix}) and disorder potential $\mat{V}$, an average expectation value over possible disorder fluctuation is obtained by following configurational average:
\begin{equation}
\begin{split}
\braket{\mat{V}}
=&\frac{1}{N} \sum_n \bra{n}\mat{V}_v\ket{n}
=\int dv D(v) \mat{V} \frac{1}{N}\sum_n\braket{n\vert n}
=\int dv D(v) \mat{V}, \\
\braket{\mat{G}}
=&\frac{1}{N}\sum_n \int dkdk'\bra{k}e^{-ikn}\mat{G} e^{ik'n}\ket{k'}
=\int dk\mat{G}(k), \\
\end{split}
\end{equation}
where local, short range scatter is assumed to satisfy $\bra{m}\mat{V}_v\ket{n}=v\delta_{m,n}$ with a probability of $D(v)$, and Bloch wavefunction $\ket{n}=\int dk e^{ikn}\ket{k}$ is used.
For multiple operators,
\begin{equation}
\begin{split}
\braket{\mat{V}\mat{G}_0\mat{V}\mat{G}_0\mat{V}}
=&\frac{1}{N}\sum_{n} \bra{n} \mat{V}\mat{G}_0\mat{V}\mat{G}_0\mat{V} \ket{n}
=\frac{1}{N^5}\sum_{n,m,l,o,p}
\bra{n}\mat{V}_v\ket{m}
\bra{m}\mat{G}_0\ket{l}
\bra{l}\mat{V}_v\ket{o}
\bra{o}\mat{G}_0\ket{p}
\bra{p}\mat{V}_v\ket{n} \\
=&\frac{1}{N^2}  \sum_{n,l}
v \bra{n}\mat{G}_0\ket{l}
v'\bra{l}\mat{G}_0\ket{n}v \\
=&\int dv_i dv_j D(v_i)D(v_j) \frac{1}{N^2}  \sum_{n,l}
v_i \bra{n}\mat{G}_0\ket{l}
v_j\bra{l}\mat{G}_0\ket{n}v_i \\
=&\int dv_i dv_j D(v_i)D(v_j)
\frac{1}{N^2} \sum_{n,l}\int dk_1dk_2dk_3dk_4
v_i \bra{k_1}e^{-ik_1n}\mat{G}_0 e^{ik_2l}\ket{k_2}
v_j\bra{k_3} e^{-ik_3l}\mat{G}_0 e^{ik_4n}\ket{k_4}v_i \\
=&\int dv_i dv_j D(v_i)D(v_j)
 \int dk_1dk_2
v_i \bra{k_1}\mat{G}_0 \ket{k_2}
v_j\bra{k_2} \mat{G}_0 \ket{k_1}v_i \\
=&\int dv_i dv_j D(v_i)D(v_j) \int dk
v_i \mat{G}_{0}(k)
v_j\mat{G}_{0}(-k) v_i, \\
\end{split}
\end{equation}
where we use $\bra{k_1}\mat{G} \ket{k_2}=\delta(k_1+k_2)G(k_1)$, due to translational invariance $G(x,x')=G(x-x')$.\footnote{$G(k,k')=\int dxdx' e^{-i(kx+k'x')}G(x,x')=\int dx'e^{-i(k+k')x'} \int dx e^{-ik(x-x')}G(x-x')=\delta(k+k')G(k)$.}

\section{Self-energy terms for first and second order in $\mat{V}$} \label{sec:VxyzCalc}

When the N\'{e}el vector is initially in $\hat{\mathbf{n}}||[100]$, we examine the self-energy induced by spin fluctuation by rewriting perturbation Hamiltonian $\mat{V}$ in Eq.~(\ref{eq:RRv}) into three components:
\begin{equation} \label{eq:Vxyz_gapless}
\begin{split}
\mat{V}_x=&-\Delta(1-\cos\theta_v)\tau_3\sigma_1, \\
\mat{V}_y=&\Delta\cos\varphi_v\sin\theta_v\tau_3\sigma_2, \\
\mat{V}_z=&\Delta\sin\varphi_v\sin\theta_v\tau_3\sigma_3.
\end{split}
\end{equation}

$\bullet$ \emph{$x$ directional fluctuation}:
Let us first consider the first order correction:
 \begin{equation} \label{eq:Vx}
 \begin{split}
\braket{\mat{V}_x}
=&-\int_{0}^{\pi} \sin\theta d\theta D_\theta \int_0^{2\pi} d\varphi D_\varphi
\Delta(1-\cos\theta)\tau_3\sigma_1
=-\Delta\int_{0}^{\pi} d\theta D_\theta   \sin\theta(1-\cos\theta)\tau_3\sigma_1, \\
\end{split}
\end{equation}
where we assume that fluctuation is white noise having an equal probability for any azimuthal angle, or $D_\varphi=\frac{1}{2\pi}$. We also assume a uniform distribution for polar angle, and define $\text{sinc}\;\theta_0=\sin\theta_0/\theta_0$. Unlike other fluctuation components, $\hat{x}$ direction has non-zero first order correction.
Furthermore, the second order correction is
 \begin{equation}
 \braket{\mat{V}_z \mat{G}_0 \mat{V}_z}
 =\braket{\mat{V}_z (\mat{G}_{0-}+\mat{G}_{0-})\mat{V}_z }.
\end{equation}
Equivalently, we compute
\begin{equation} \label{eq:VxGVx}
\begin{split}
 \braket{ \mat{V}_x\mat{G}_{0\pm}\mat{V}_x }
 =&
  \int_{0}^{\pi} \sin\theta  d\theta D_\theta \int_0^{2\pi} d\varphi D_\varphi
 \Delta^2 (1-\cos\theta)^2
\int_{BZ} \frac{d^3k}{(2\pi)^3} \frac{1}{\epsilon-(a_0\pm a)+i0^+}\frac{1}{2} \\
 &\times
 \tau_3\sigma_1\left(
 1\pm \frac{1}{a}(a_1\tau_1+a_2\tau_2+ a_3\tau_3\sigma_1+a_4\tau_3\sigma_2)
 \right)_k \tau_3\sigma_1 \\
 =&
 \left( A_{0\pm}
 \mp A_{1\pm} \tau_1
 \pm  A_{3\pm} \tau_3\sigma_1
 \right)  \Delta^2\int_{0}^{\pi} d\theta D_\theta \sin\theta(1-\cos\theta)^2,  \\
 \end{split}
 \end{equation}
 where we integrate the Green function over the whole Brillouin zone for the reason explained in Appendix~\ref{sec:evaluation}. Note that $a$ is not even nor odd function in $k_y$ due to $a_3=-(\lambda-\lambda_z\cos k_z)\sin k_y+\Delta$ term. Nevertheless, $a$ is an even function in $k_x$ and $k_z$ and, therefore, $a_2$ and $a_4$ have a vanishing magnitude in Eq.~(\ref{eq:VxGVx}) due to the fact that they are odd in $k_z$ and $k_x$, respectively. The remaining terms in Eq.~(\ref{eq:VxGVx}) are
 \begin{equation}
\begin{split}
A_{0\pm}=&\int_{BZ} \frac{d^3k}{(2\pi)^3}
 \frac{1}{\epsilon-(a_0\pm a)+i0^+}\frac{1}{2}, \\
A_{1\pm}=& \int_{BZ} \frac{d^3k}{(2\pi)^3}
 \frac{1}{\epsilon-(a_0\pm a)+i0^+}\frac{1}{2}\frac{a_1}{a}, \\
A_{3\pm}=& \int_{BZ} \frac{d^3k}{(2\pi)^3}
 \frac{1}{\epsilon-(a_0\pm a)+i0^+}\frac{1}{2}\frac{a_3}{a}.\\
\end{split}
\end{equation}

$\bullet$ \emph{$y$ directional fluctuation}: Let us now consider $\mat{V}_y$. The first order correction is
 \begin{equation}
\braket{\mat{V}_y}
=\int_{0}^{\pi} \sin\theta  d\theta D_\theta \int_0^{2\pi} d\varphi D_\varphi
\Delta \cos\varphi\sin\theta\tau_3\sigma_2
=0,
\end{equation}
where we assume an equal probability for any azimuthal angle by setting $D_\varphi=\frac{1}{2\pi}$.
For the second order correction, we compute
\begin{equation} \label{eq:VyGVy}
\begin{split}
 \braket{ \mat{V}_y\mat{G}_{0\pm}\mat{V}_y }
 =&
  \int_{0}^{\pi} \sin\theta  d\theta D_\theta \int_0^{2\pi} d\varphi D_\varphi
 \Delta^2 \cos^2\varphi\sin^2\theta
  \int_{BZ} \frac{d^3k}{(2\pi)^3}
 \frac{1}{\epsilon-(a_0\pm a)+i0^+}\frac{1}{2}\\
 &\times
 \tau_3\sigma_2\left(
 1\pm \frac{1}{a}(a_1\tau_1+a_2\tau_2+ a_3\tau_3\sigma_1+a_4\tau_3\sigma_2)
 \right)_k \tau_3\sigma_2 \\
 =&
 \left( A_{0\pm}
 \mp A_{1\pm} \tau_1
 \mp  A_{3\pm} \tau_3\sigma_1
 \right)  \frac{\Delta^2}{2}\int_{0}^{\pi} d\theta D_\theta \sin^3\theta.
 \\
 \end{split}
 \end{equation}

 $\bullet$ \emph{$z$ directional fluctuation}: Lastly, we consider $\mat{V}_z$. The first order correction is
\begin{equation}
\braket{\mat{V}_z}
=\int_{0}^{\pi}\sin\theta d\theta D_\theta \int_0^{2\pi} d\varphi D_\varphi
\Delta \sin\varphi\sin\theta\tau_3\sigma_3
=0,
\end{equation}
where we assume $D_\varphi=\frac{1}{2\pi}$. Furthermore, the second order correction is
\begin{equation} \label{eq:VzGVz}
\begin{split}
 \braket{ \mat{V}_z\mat{G}_{0\pm}\mat{V}_z }
 =&
  \int_{0}^{\pi}\sin\theta d\theta D_\theta \int_0^{2\pi} d\varphi D_\varphi
 \Delta^2 \sin^2\varphi\sin^2\theta
 \int_{BZ} \frac{d^3k}{(2\pi)^3}
 \frac{1}{\epsilon-(a_0\pm a)+i0^+}\frac{1}{2}\\
 &\times
  \tau_3\sigma_3 \left(
 1\pm \frac{1}{a}(a_1\tau_1+a_2\tau_2+ a_3\tau_3\sigma_1+a_4\tau_3\sigma_2)
 \right)_k  \tau_3\sigma_3\\
 =&
 \left( A_{0\pm}
 \mp A_{1\pm} \tau_1
 \mp  A_{3\pm} \tau_3\sigma_1
 \right)  \frac{\Delta^2}{2}\int_{0}^{\pi} d\theta D_\theta \sin^3\theta.
 \\
 \end{split}
 \end{equation}

In summary, the self-averaged self-energy term in Eq.~(\ref{eq:Sigeff2}) becomes
\begin{equation} \label{eq:Sigeff_final}
\begin{split}
\mat{\Sigma}_{eff}=& \braket{\mat{V}} + \braket{\mat{V}\mat{G}_0\mat{V}} \\
=&  \braket{\mat{V}_x} + \braket{\mat{V}_y} + \braket{\mat{V}_z}
+\braket{(\mat{V}_x+\mat{V}_y+\mat{V}_z)\mat{G}_0(\mat{V}_x+\mat{V}_y+\mat{V}_z)} \\
=&
\braket{\mat{V}_x} + \braket{\mat{V}_y} + \braket{\mat{V}_z}
+\braket{\mat{V}_x\mat{G}_0\mat{V}_x}
+\braket{\mat{V}_y\mat{G}_0\mat{V}_y}
+\braket{\mat{V}_z\mat{G}_0\mat{V}_z} \\
=&\Sigma_0
+\Sigma_1\tau_1
+\Sigma_3\tau_3\sigma_1,
\end{split}
\end{equation}
where we utilize the fact that $\braket{ \mat{V}_i\mat{G}_{0}\mat{V}_j}=0$ for $i\neq j$. By combining the results in Eqs.~(\ref{eq:Vx}-\ref{eq:VzGVz}), we obtain $\Sigma_0$, $\Sigma_1$, and $\Sigma_3$ terms up to the second order in $\mat{V}$ among possible self-energies $\Sigma_{0,1,2,3,4,5}$ in Eq.~(\ref{eq:Sigeff2}). The resultant self-energy terms for $\Sigma_{0,1,3}$ are summarized in Eq.~(\ref{eq:Sigma_gapless}).

 \end{widetext}

\bibliography{reference}		

\end{document}